\documentclass[%
 reprint,
 amsmath,amssymb,
 aps,
]{revtex4-2}

\usepackage{graphicx}
\usepackage{dcolumn}
\usepackage{bm}
\usepackage{amsmath}
\usepackage{amssymb}
\usepackage{txfonts}
\usepackage{mathrsfs}
\usepackage{soul, xcolor}
\usepackage[mathlines]{lineno}

\def\b0{{\mathbf 0}}

\def\b0{{\mathbf 0}}

\def\beq{\begin{equation}}
\def\eeq{\end{equation}}

\begin{document}

\title{Generalized Hertz action and quantum criticality of two-dimensional Fermi systems}
\author{Mateusz Homenda  }
\email{mateusz.homenda@fuw.edu.pl}
\affiliation{Institute  of Theoretical Physics, Faculty of Physics, University of Warsaw, Pasteura 5, 02-093 Warsaw, Poland}
\author{Pawel Jakubczyk } 
\email{pawel.jakubczyk@fuw.edu.pl}
\affiliation{Institute  of Theoretical Physics, Faculty of Physics, University of Warsaw, Pasteura 5, 02-093 Warsaw, Poland} 
\author{Hiroyuki Yamase } 
\email{yamase.hiroyuki@nims.go.jp}
\affiliation{Research Center for Materials Nanoarchitectonics, National Institute for Materials Science, Tsukuba 305-0047, Japan}

\date{\today}

\begin{abstract} 
We reassess the structure of the effective action and quantum critical singularities of two-dimensional Fermi systems characterized by the ordering wavevector $\vec{Q}= \vec{0}$. By employing  infrared cutoffs on all the massless degrees of freedom, we derive a generalized form of the Hertz action, which does not suffer from problems of singular effective interactions. We demonstrate that the  Wilsonian momentum-shell  renormalization group (RG) theory capturing the infrared scaling should be formulated keeping $\vec{Q}$ as a flowing, scale-dependent quantity. At the quantum critical point, scaling controlled by the dynamical exponent $z=3$ is overshadowed by a broad scaling regime characterized by a lower value of $z \approx 2$. This in particular offers an explanation of the results of quantum Monte Carlo simulations pertinent to the electronic nematic quantum critical point.  
\end{abstract}

\maketitle

\emph{Introduction} 
Quantum criticality in Fermi systems constitutes a highly relevant and largely open problem for condensed matter theory. Its significance stems from the growing experimental evidence  demonstrating non-Fermi-liquid behavior of thermodynamic as well as transport properties at the onset of different ordered states in a diversity of compounds; the high-$T_c$ cuprate superconductors being the most prominent examples\cite{Keimer_2015}.
A persistent question concerns the structure of the low-energy effective  action to correctly capture the critical singularities at quantum criticality in Fermi systems. The traditional approach, first developed by Hertz\cite{Hertz_1976} and later extended by Millis\cite{Millis_1993}, borrowed the spirit of the Wilsonian theory of  classical critical phenomena\cite{Cardy_1996, Goldenfeld_1992}. It proposed to  integrate out the original  degrees of freedom resulting in an exact representation of the problem in terms of an effective order parameter action. In the subsequent step this action was expanded in powers of the ordering field, truncating at quartic order; the two-point function was replaced by its low momentum/frequency asymptotic form, and  the (supposedly irrelevant) momentum/frequency structure of the bosonic self-interaction was  disregarded. This leads to a relatively simple Hertz action\cite{Hertz_1976, Millis_1993, Nagaosa_1998} 
\begin{equation} 
\label{H_action}
S_{H}[\phi]=\int_q \phi_{-\vec{q},-q_0}\left[m^2 + Z \vec{q}^2+ A\frac{|q_0|}{|\vec{q}|} \right]   \phi_{\vec{q},q_0} +u\int_x\phi(x)^4\;
\end{equation}
describing the propagation of a damped collective bosonic mode $\phi$, where the interaction with fermions is described by the so-called Landau damping term $\sim |q_0|/|\vec{q}|$, $q_0$ being the frequency and $\vec{q}$ the momentum of the order parameter field.
Here $\{m^2, Z, A, u\}$ are constants, $q:=(q_0,\vec{q})$, $\int_q:=1/(2\pi)^3\int d q_0 \int d^2q$, and $\int_x:=\int d\tau \int d^2 x$ encompasses integration over space and the imaginary time $\tau$.
The form $\sim |q_0|/|\vec{q}|$ is valid for instabilities occurring at ordering wavevector $\vec{Q}=\vec{0}$. 

 In contrast to classical statistical physics systems, this procedure involves integrating out gapless particle-hole excitations across the Fermi surface, the consequence of which becomes revealed by inspection of the nature of the frequency/momentum expansion of the bosonic interaction vertices (for example the fermionic box diagram), which turns out to be singular at $T = 0$ \cite{Belitz_2005, Lohneysen_2007, Metlitski_1010, Thier_2011}. For this reason quantum critical Fermi systems (at least in dimensionality $d=2$  and temperature $T=0$) cannot be adequately described by a purely bosonic action characterized by local interactions. 

The above issues motivated development of a diversity of approaches that retain the fermionic degrees of freedom, which are coupled to order parameter fluctuations\cite{Lee_2009, Maslow_2010, Metlitski_1010, Mross_2010, Drukier_2012, Fitzpatrick_2013, Dalidovich_2013, Mandal_2015, Holder_2015, Ridgway_2015, Punk_2016, Trott_2018, Lee_2018, Damia_2019, Damia_2020, Saterskog_2021, Zhang_2023, Mayrhofer_2024}.  These  theoretical routes come with their own questions. One of these concerns the transition between the microscopic and effective low-energy action. This is transparent, for instance, in the analysis concerning generation of the damped dynamics of the bosonic mode. As was emphasized in previous literature (see in particular Ref.~\onlinecite{Fitzpatrick_2013}), appearance of the $\sim |q_0|/|\vec{q}|$ term requires that the fermions be integrated out down to the Fermi level. 
It is not conceivable to generate the standard Landau damping term by a Wilsonian-type RG flow until the cutoff on fermions is completely removed and therefore the fermionic degrees of freedom become once and for all integrated out of the theory. In consequence, accounting for the Landau damping (in its standard form) within such approaches requires fully dressing the boson propagator with self-energy before calculating any loops that involve internal boson lines in the coupled Bose-Fermi theory. 

In the present work we systematically readdress the theory of quantum criticality in Fermi systems featuring $\vec{Q}=\vec{0}$ instabilities  and develop a Wilsonian RG approach, where both the bosonic and fermionic propagators become equipped with momentum cutoffs, and, upon lowering these, generate the RG flow, leading to a generalization of the Hertz action.
Our goal is to develop an approximate approach fully encompassing the Hertz-Millis framework, but at the same time refraining from completely integrating femions out, such that the singular effective bosonic interactions never appear. 
We  demonstrate that the RG flow of the bosonic properties involves a completely different,  not previously recognized contribution, which encodes a structure  richer than the conventional Hertz-Millis theory. 

\begin{figure}
\includegraphics[width=7.5cm]{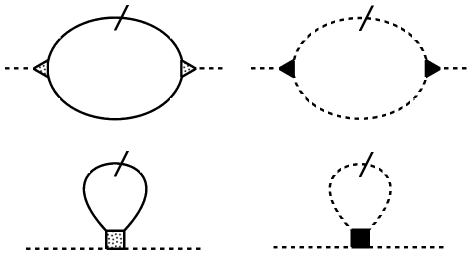}
\caption{Terms contributing to the flow of the bosonic 2-point function. Dressed (scale dependent) fermion and boson propagators are depicted as full and dashed lines respectively. Black triangles and rectangles represent the bosonic vertices, while dotted (grey) triagles and rectangles stand for fermion-boson interactions. The stroked lines represent the single-scale propagators: $\mathcal{S}_{f}$ for fermion and $\mathcal{S}_{b}$ for boson propagators (see the main text).} 
\label{Diagrams_1}
\end{figure}

\emph{Generalized Hertz action} 
Our approach relies on the nonperturbative RG framework in the Wetterich formulation\cite{Wetterich_1993, Kopietz_book, Dupuis_2021}. 
This methodology has, in recent years, led to several important new insights concerning key problems of condensed matter theory, critical systems in particular. Examples include identification of strong-coupling fixed points for the Kardar-Parisi-Zhang problem in $d>1$ \cite{Canet_2010}, resolution of the problem of dimensional reduction and its breaking for the random field Ising model \cite{Tissier_2011}, discovery of new multicritical RG fixed points for the $O(N)$ models in $d=3$ \cite{Yabunaka_2017}, and invalidation \cite{Chlebicki_2021} of the predictions of perturbative approaches concerning nonanalyticity of the critical exponents as function of $d$ and $N$.   
In our approach we integrate the coupled fermionic ( $\{\bar{\psi}, \psi \}$ ) and bosonic ( $\phi$ ) fluctuating fields out of the partition function  via a renormalization group flow. The central object  is the scale-dependent effective action $\Gamma^\Lambda[\bar{\psi},\psi,\phi]$, which continuously interpolates between the bare effective action $S[\bar{\psi},\psi,\phi]$  and the full effective action (free energy), when the infrared cutoff $\Lambda$ is lowered from the $UV$ scale towards zero. Below we suppress the arguments of $\Gamma$ for readability. The evolution of $\Gamma$ upon varying $\Lambda$ is governed by the exact Wetterich flow equation 
\begin{equation}
\label{Wetterich_eq}
\dot{\Gamma}=\beta_b+\beta_f\;,
\end{equation}
where 
\begin{equation}
\label{beta_b}
\beta_b = \frac{1}{2} {\rm Tr}\left\{  \dot{\mathcal{R}}_b
(\tilde{\Gamma}^{(2)}_{\phi\phi})^{-1}\left[1-\tilde{\Gamma}^{(2)}_{\phi\psi} (\tilde{\Gamma}^{(2)}_{\psi\psi})^{-1}  \tilde{\Gamma}^{(2)}_{\psi\phi} (\tilde{\Gamma}^{(2)}_{\phi\phi})^{-1} \right]^{-1}    \right\}\; ,
\end{equation}
\begin{equation}
\label{beta_f}
\beta_f = \frac{1}{2} {\rm Tr}\left\{\dot{\mathcal{R}}_f(\tilde{\Gamma}^{(2)}_{\psi\psi})^{-1}\left[1-\tilde{\Gamma}^{(2)}_{\psi\phi} (\tilde{\Gamma}^{(2)}_{\phi\phi})^{-1}  \tilde{\Gamma}^{(2)}_{\phi\psi} (\tilde{\Gamma}^{(2)}_{\psi\psi})^{-1} \right]^{-1}    \right\}\;. 
\end{equation} 
The quantity $\tilde{\Gamma} := \Gamma +\Delta S$ denotes the action $\Gamma$ supplemented with the regulator term $\Delta S =\frac{1}{2}\Phi ( \mathcal{R} \Phi^T )$, which is quadratic in the fields $\Phi=(\bar{\psi}, \psi, \phi)$ and contains bosonic ($\mathcal{R}_b$) and fermonic ($\mathcal{R}_f$) components. The quantity $\tilde{\Gamma}^{(2)}$ denotes the second (functional) field derivative of $\tilde{\Gamma}$ with the relevant fields specified by the subscript in each case. By $\dot{X}$ we mean $\partial_\Lambda X$. Finally, the trace (Tr) sums over the field components, momenta and frequencies. Our notation is equivalent to that introduced in Ref.~\onlinecite{Obert_2013} (for details and derivations see also Ref.~\onlinecite{Obert_2014_PhD}) with the exception that $\phi$ is a real scalar in our case. 
Differentiating the flow equation [Eq.~(\ref{Wetterich_eq})]  with respect to fields gives rise to an hierarchy of flow equations for the one-particle irreducible vertex functions. We concentrate on the RG flow equation for the bosonic two-point function, obtained by taking the  second functional derivative of Eq.~(\ref{Wetterich_eq}) with respect to $\phi$.  The resulting equation \cite{Kopietz_book, Obert_2013, Obert_2014_PhD, Dupuis_2021} involves terms represented via one-loop Feynman diagrams  depicted in Fig.~\ref{Diagrams_1}. 

The expressions involve the flowing fermion propagator $\tilde{G}:=(\tilde{\Gamma}^{(2)}_{\psi\psi})^{-1}$ supplemented with a momentum cutoff $R_f^\Lambda(\vec{k})$:
\begin{equation}
\tilde{G}_{k,k',\sigma,\sigma'}=\left(-i k_0+\xi_{\vec{k}}+R_f^\Lambda(\vec{k})+\Sigma^\Lambda(k)\right)^{-1}\delta_{k,k'}\delta_{\sigma,\sigma'}\; 
\label{G_full}
\end{equation}
[with $k:=(k_0,\vec{k})$];  the flowing regularized boson propagator $\tilde{G_b}:=(\tilde{\Gamma}^{(2)}_{\phi\phi})^{-1}$,  interaction vertices, as well as the so-called single-scale propagators defined as 
\begin{align} 
\mathcal{S}_{f}:=-(\tilde{\Gamma}^{(2)}_{\psi\psi})^{-1} \dot{\mathcal{R}}_f (\tilde{\Gamma}^{(2)}_{\psi\psi})^{-1} \\
\mathcal{S}_{b}:=-(\tilde{\Gamma}^{(2)}_{\phi\phi})^{-1}\dot{\mathcal{R}}_b(\tilde{\Gamma}^{(2)}_{\phi\phi})^{-1} \; . 
\end{align}
 In Fig.~\ref{Diagrams_1} the single-scale propagators correspond to stroked lines. The bare (microscopic) action contains only contributions quadratic in fields and a Yukawa-type term coupling the bosonic field with two fermionic variables \cite{Nagaosa_1998}. In addition, the bare boson propagator carries no momentum/frequency dependence. These dependencies are  generated by gradually integrating the fermions out via the contribution to the flow given by the first diagram in Fig.~\ref{Diagrams_1}. 

\begin{figure}
\includegraphics[width=7.5cm]{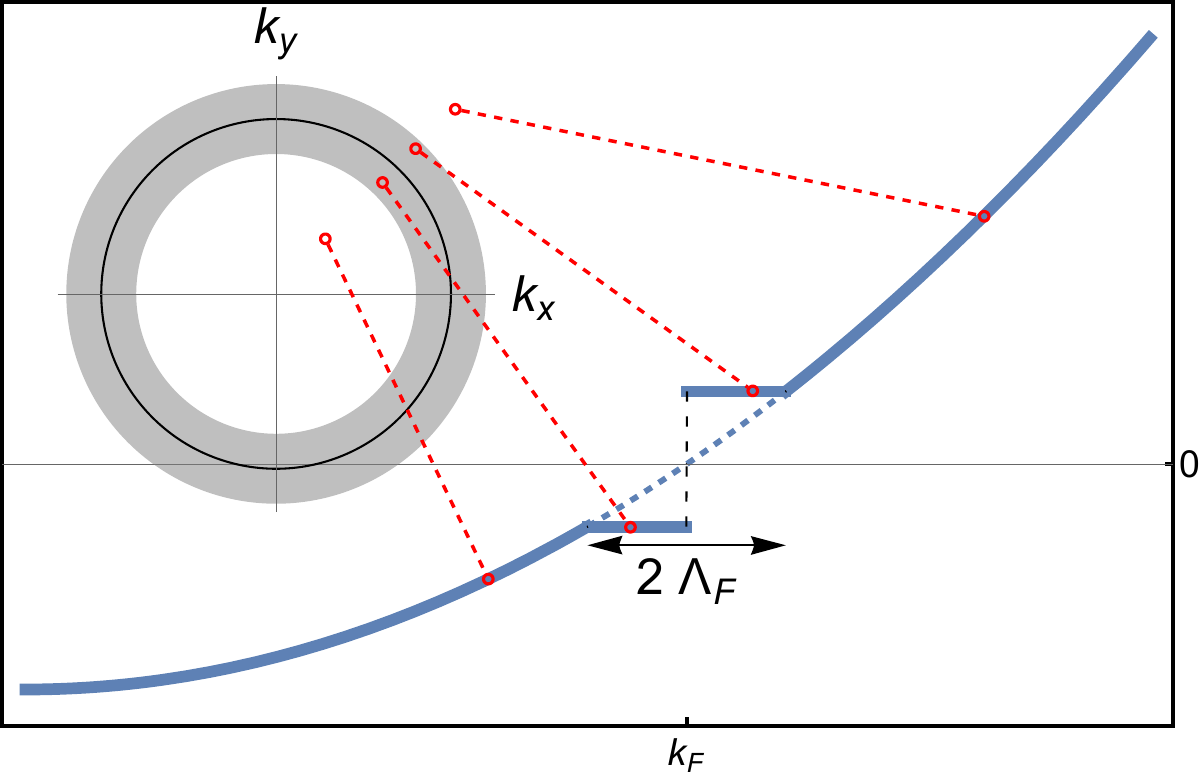}
\caption{A schematic plot of the regularized dispersion $\xi_{\vec{k}}+R_f(\vec{k})$. Including the regulator introduces a deformation of the dispersion in a strip of extension $2\Lambda_F(\Lambda)$ around the Fermi level. In the inset the black line represents the Fermi surface and gray shell designates the area of the deformation. } 
\label{Disp}
\end{figure}

 The flow parameter  $\Lambda$ appearing in the Wetterich equation is identified with the bosonic momentum cutoff ($\Lambda_b=\Lambda$). The precise form of the bosonic cutoff will be specified later. We will use the following form of the cutoff function on fermions:  
\begin{align}
 R_f(\vec{k})=\begin{cases} \left(\xi_{k_F+\Lambda_F}-\xi_{\vec{k}}\right)\theta \left(\Lambda_F-(|\vec{k}|-k_F)\right)\;\; {\rm for}\;\; |\vec{k}|\geq k_F \\ 
 \left(\xi_{k_F-\Lambda_F}-\xi_{\vec{k}}\right)\theta \left(\Lambda_F-(k_F-|\vec{k}|)\right)\;\; {\rm for}\;\; |\vec{k}|< k_F\;.
 \end{cases}
\end{align}
The quantity $\Lambda_F=\Lambda_F(\Lambda)$ is a function of $\Lambda$. The effect of adding $R_f(\vec{k})$ to the dispersion $\xi_{\vec{k}}$ amounts to deforming it in a sliver of extension $2\Lambda_F$ around the  Fermi level, as depicted in Fig.~\ref{Disp}. 
We expect that our conclusions are completely insensitive to the precise choice of the momentum cutoff function $R_f(\vec{k})$.

The Hertz-like approach corresponds in our framework to sending $\Lambda_F$ to zero before $\Lambda$. This can be realized e.g. by taking 
\begin{equation}
\Lambda_F=(\Lambda-\Lambda_0)\theta(\Lambda-\Lambda_0) 
\label{LamF}
\end{equation}
with $\Lambda_0>0$, such that $\Lambda_F$ becomes zero at positive $\Lambda$. In what follows, we will perform a detailed comparison between the pictures emergent for $\Lambda_0=0$ and $\Lambda_0>0$ 
(Hertz-Millis case).

The flow equation represented by the terms depicted in Fig.~\ref{Diagrams_1} is exact, but can be solved only approximately. Its present truncation is devised such that it encompasses the Hertz-Millis theory if $\Lambda_F$ is scaled to zero first [e.g when one takes $\Lambda_0>0$ in Eq.~(\ref{LamF})], but does not require this in any way. The gradual generation of the dynamics of the boson propagator can be followed upon reducing $\Lambda$ towards zero. The key present approximation amounts to disregarding the Fermi self energy [$\Sigma^\Lambda(k)=0$] and the flow of the Yukawa coupling $g$ as well as other fermionic interactions generated by the flow. This allows us to write the contribution to the flow of the boson propagator represented by the first diagram in Fig.~\ref{Diagrams_1} as: 
\begin{align}
\mathcal{X}(q,\Lambda_F)=-2g^2\int_k \partial_\Lambda R_f(\vec{k})\tilde{G}_0(k)^2\left(\tilde{G}_0(k+q)+\tilde{G}_0(k-q)\right)\;,    
\label{bubble_reg}
\end{align} 
where the scale-dependent (regularized) fermion propagator is given by
$
\tilde{G}_0(k)^{-1}=[-ik_0+\xi_{\vec{k}}+R_f(\vec{k})].
$
To simplify the calculations and highlight the new theoritical insight clearly, we employ the standard quadratic dispersion $\xi_{\vec{k}}=(\vec{k}^2-k_F^2)/2m_f$.  We then evaluate the integrals in Eq.~(\ref{bubble_reg}) and subsequently integrate over the cutoff scale, which results in the frequency/momentum structure of the boson propagator (generated from integrating the fermions from the UV cutoff scale $\Lambda_u$ down to the scale $\Lambda$). Computing 
\begin{align} 
B\left(\vec{q},q_0,\Lambda_F(\Lambda)\right):=\int_{\Lambda_u}^{\Lambda} d\Lambda' \mathcal{X}\;,
\label{X_int}
\end{align} 
we obtain: 
\begin{align} 
\label{B_final}
B(\vec{q},q_0,\Lambda_F)= B_< \theta( -|\vec{q}|+\Lambda_F)+ B_> \theta(|\vec{q}|-\Lambda_F)\;,
\end{align} 
where 
\begin{align}
\label{B_m}
  B_< =  -\mathcal{N}_<\frac{|\vec{q}|\Lambda_F}{q_0^2+4v_F^2\Lambda_F^2}\;,
\end{align} 
\begin{align} 
\label{B_w}
B_> \approx -\mathcal{N}_<\frac{\vec{q}^2}{q_0^2+4v_F^2\vec{q}^2} 
+ \mathcal{N}_> \frac{q_0}{|\vec{q}|}\left[\arctan\frac{2v_F|\vec{q}|}{q_0}-\arctan\frac{2v_F\Lambda_F}{q_0}\right],    
\end{align}
and $\mathcal{N}_<=\mathcal{N}_> v_F^3=4g^2k_Fv_F/\pi^2$. The above expressions are essential for the present work. Eq.~(\ref{B_m}) follows from exactly evaluating the integrals of Eq.~(\ref{bubble_reg}) for $|\vec{q}|<\Lambda_F$ and subsequently integrating over the cutoff scale according to Eq.~(\ref{X_int}). Eq.~(\ref{B_w}) results from evaluating Eq.~(\ref{bubble_reg}) for $|\vec{q}|\gg\Lambda_F$ retaining the terms, which generate the standard Landau damping $\sim |q_0|/|\vec{q}|$ if we first take $\Lambda_F\to 0$ and subsequently consider $\frac{|\vec{q}|}{q_0}\to \pm\infty$; the dropped terms are regular in $q$ in the limit $\Lambda_F\to 0$ and we made no assumptions concerning the relative magnitude of $|q_0|$ and $v_F|\vec{q}|$. Concerning the structure of $B(\vec{q},q_0,\Lambda_F)$, we emphasize that: $(i)$ for $\Lambda_F\to 0$ it recovers, via $B_>$, the standard Landau damping term of the Hertz action; $(ii)$ it takes minimum at $(q_0,|\vec{q}|)=(0,\Lambda_F)$, which indicate that the ordering wavevector depends on the cutoff scale and falls at $|\vec{Q_\Lambda}| = Q_\Lambda=\Lambda_F$, thus scaling to zero under RG. Note in particular that artificially putting $Q_\Lambda=0$ suppresses the flow of the mass generated from fermionic bubble. This explains (and evades) the unwelcome features of the mass flow under Wilsonian RG, discussed in Ref.~\onlinecite{Fitzpatrick_2013}. Observe that the mass flow is generated from $B_<$ [evaluated at $(q_0,|\vec{q}|)=(0,\Lambda_F)$]. 
In the present generalization of the Hertz-Millis approach we will parametrize the flowing inverse boson propagator as 
\begin{equation}
\Gamma_\Lambda^{(2)}= Z(|\vec{q}|-Q_\Lambda)^2+A q_0^2 +m_\Lambda^2+B(\vec{q},q_0,\Lambda_F(\Lambda))\;.
\label{H_M_gen}
\end{equation}    
The essential modification of the standard Hertz action amounts to replacing the term $\sim |q_0|/|\vec{q}|$ occurring in Eq.~(\ref{H_action}) with the formula $B(\vec{q},q_0,\Lambda_F)$ obtained above, such that fermionic fluctuations are included only down to the scale $\Lambda_F(\Lambda)$ (which is sent to zero as $\Lambda\to 0$). We emphasize that the RG flow of the boson propagator will be strongly influenced by the first term in Eq.~(\ref{B_final}), corresponding to $|\vec{q}|$ small. 

\begin{figure}[h]
    \includegraphics[width=9.5cm]{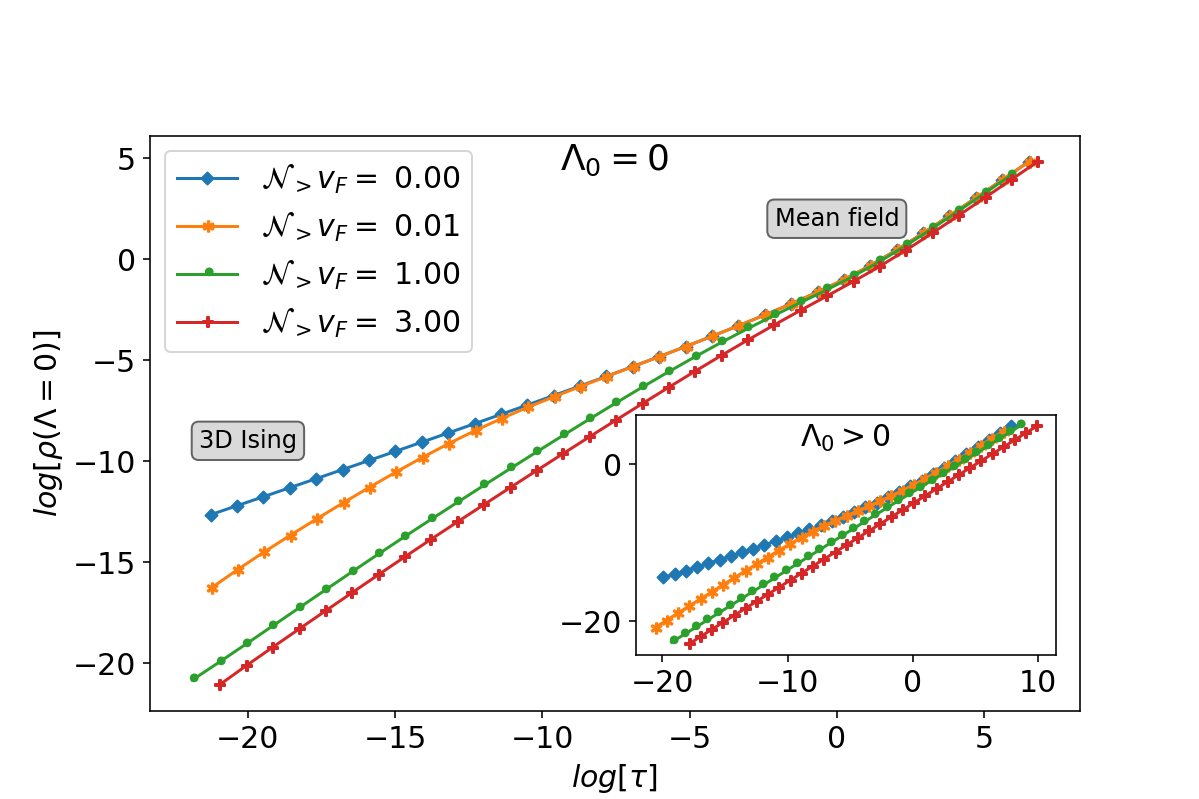}
    \caption{The renormalized value of $\rho=\phi_0^2/2$ plotted as function of the control parameter $\tau$ (bare mass) at $T=0$ for a sequence of values of the boson-fermion coupling $g$ ( $\mathcal{N}_> v_F \propto g^2$). The uppermost curve (blue) exhibits crossover between scaling pertinent to the classical 3d Ising universality class and mean-field scaling outside the true critical region. Upon gradually switching on $g$ the asymptotic critical scaling corresponding to efffective dimensionality $D=d+z\geq 4$ sets in and the system again crosses over to mean-field behavior. The behavior exhibited in the inset corresponds to $\Lambda_0>0$
    (integrating out fermions first).  There is no qualitative difference between these two cases, 
    which demonstrates that including $B_<$ has no impact on the critical singularities of the order parameter. }
    \label{phase_diag}
\end{figure}

\emph{The dynamical exponent}
We now examine the consequences of the term $B_<$ for the dynamical exponent $z$ of the order parameter field. If we first take $\Lambda_F\to 0$ 
setting $\Lambda_0 \neq 0$ in Eq.~(\ref{LamF})
[thus removing the term $B_<$ from $B$ in Eq.~(\ref{H_M_gen})], and subsequently consider the limit $v_F|\vec{q}|/q_0\to\pm\infty$, we recover the Hertz result $z=z_{H}=3$, proceeding along the standard path \cite{Nagaosa_1998}. 


The situation radically changes, if we instead integrate both bosons and fermions in parallel by considering Eq.~(\ref{LamF}) with $\Lambda_0=0$ 
($\Lambda_f = \Lambda$ )
, in which case $B_<$ plays a prominent role. The anticipated value of $z$ resulting from the $q_0$ dependence of $B_<$ can be deduced by putting $|\vec{q}|=\Lambda$ in $B_<$ and expanding for $q_0\ll 2 v_F \Lambda$. 
The leading term renormalizes the mass $m_{\Lambda}^2$ in Eq.~(\ref{H_M_gen}) and the second term is proportional to $q_0^2 / \Lambda^2$.
We find that  the $B_<$ term in $\Gamma_\Lambda^{(2)}$ scales as $\Lambda^2$ (thus leading to scale invariant propagator) provided $q_0\sim |\vec{q}|^2$ which corresponds to the dynamical exponent $z=z_<=2$. We also note that the choice $\Lambda_F\sim \Lambda$ is the only one, which allows to write $\Gamma _\Lambda^{(2)}$ given by Eq.~(\ref{H_M_gen}) (keeping either $B_<$ or $B_>$) in a scaling form. From this above heuristic picture one anticipates a competition between two scaling behaviors governed by $z\approx 2$ and $z\approx 3$. This is checked and confirmed by solving the RG equations as described below.


\begin{figure}
\includegraphics[width=9.5cm]{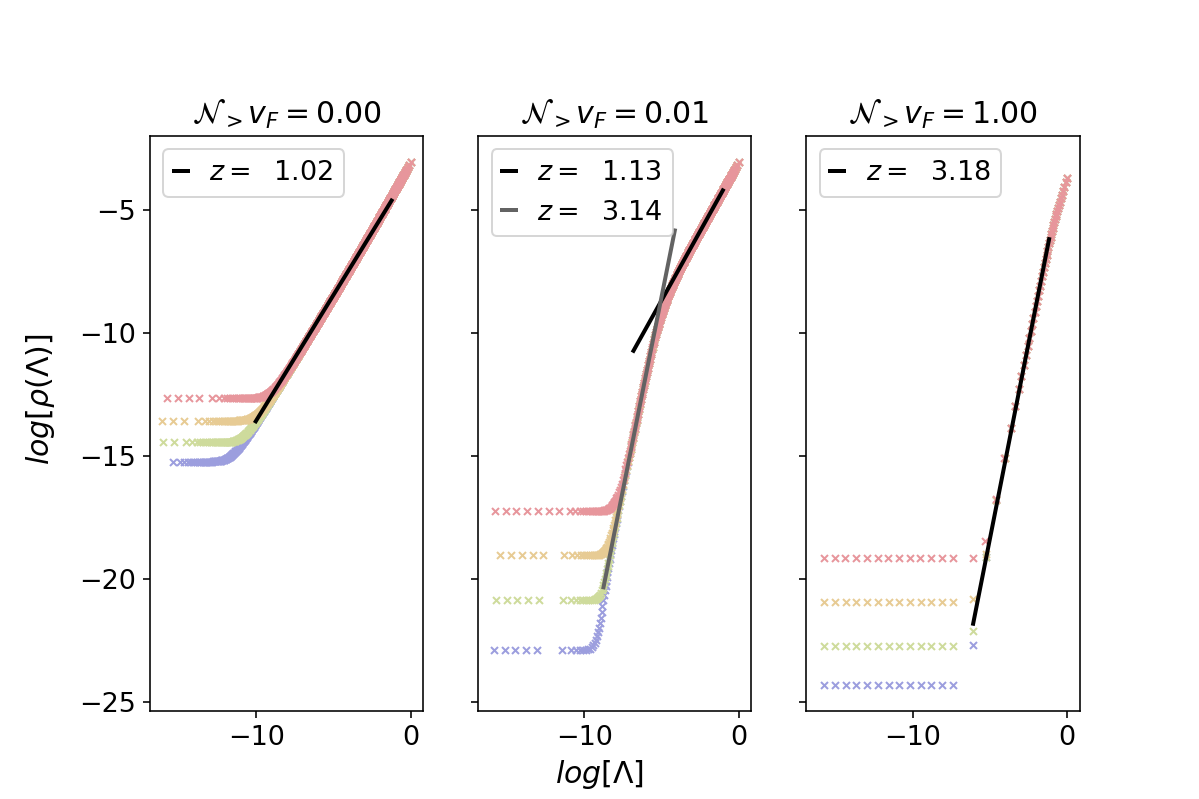}
\caption{ The RG flows of $\rho^\Lambda=(\phi_0^\Lambda)^2/2$ for a sequence of values of $\tau$ progressively tuning the system towards the QCP in a situation, where fermions are integrated first [$\Lambda_0>0$ in Eq.~(\ref{LamF})]. The value of $z$ can be read off by fitting the power law (see the main text). The crossover between $z=1$ and  $z=3$ is clearly visible both as function of $g$ ( $\mathcal{N}_> v_F \propto g^2$) and the cutoff scale $\Lambda$.  } 
\label{HM_z_plots}
\end{figure}

\emph{RG flow} 
A convenient way to extract the dynamical exponent $z$ from the RG flow, capturing possible crossovers, is to  inspect the behavior of the flowing order parameter expectation value, which follows $\phi_0^\Lambda \sim\Lambda^{z/2}$. This is, for $d=2$, implied from $m_\Lambda^2\sim\Lambda^2$, $u^\Lambda \sim \Lambda^{4-(d+z)}$ and the relation $m_\Lambda^2=u^\Lambda(\phi_0^\Lambda)^2$ (see e.g. Refs.~\onlinecite{Berges_2002, Jakubczyk_2008}). Equivalently, one may invoke the scaling dimension of the $\phi$  field $[\phi]=d+z-2+\eta$, which gives $z/2$ for $d=2$. Here we neglect the anomalous dimension $\eta$. 

We evaluate the flow of the boson order parameter $\phi_0^\Lambda$ and quartic coupling $u^\Lambda$ within a simple truncation of the Wetterich equation, where the bosonic propagator is dressed as dictated by Eq.~({\ref{H_M_gen}). We include\cite{Jakubczyk_2008, Bauer_2011} the renormalization of $u$ via bosonic fluctuations of order $\sim u^2$, which allows for capturing also the 3d Wilson-Fisher fixed point. The latter governs the critical behavior in the absence of the Fermi-Bose coupling $g$ and gives rise to an intermediate scaling regime described by the dynamical exponent $z=1$, as observed in Ref.~\onlinecite{Fitzpatrick_2013} and also clearly captured in our approach (see Fig.~\ref{phase_diag}). Within our present framework, the flow equations for $\phi_0^\Lambda$ and the quartic coupling $u^\Lambda$ are derived following the standard procedure described for example in Refs.~\onlinecite{Berges_2002, Jakubczyk_2008}. We choose the Litim cutoff\cite{Litim_2001} on bosons  
\begin{equation}
R_b(\vec{q})=Z[\Lambda^2-(|\vec{q}|-Q_\Lambda)^2]\theta [\Lambda^2-(|\vec{q}|-Q_\Lambda)^2]    \;.
\end{equation}
We verified, that implementing the Wetterich cutoff\cite{Berges_2002} instead, does not change any of our results. Our major conclusion concerning $z$ is best summarized by comparing the RG flows of the order parameter depicted in Fig.~\ref{HM_z_plots} and Fig.~\ref{l0_0_z_plots}. 


\begin{figure}
\includegraphics[width=9.5cm]{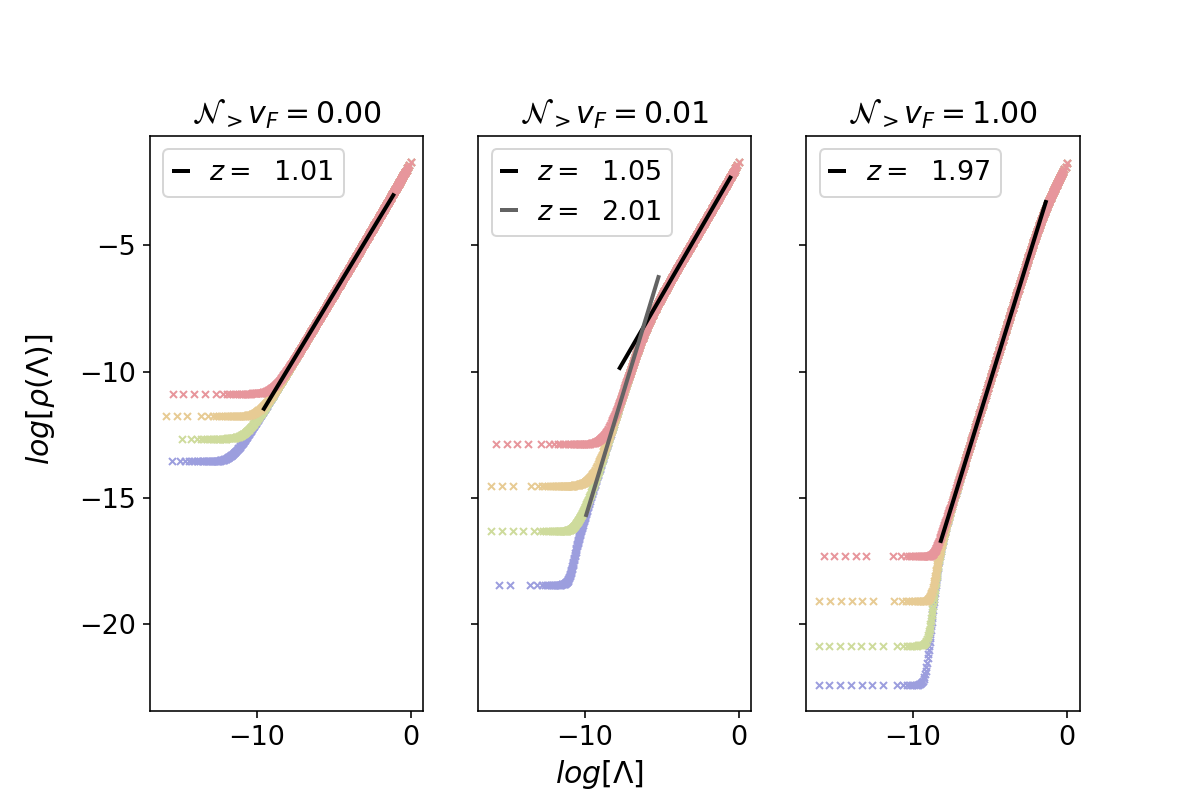}
\caption{ The RG flows of $\rho^\Lambda=(\phi_0^\Lambda)^2/2$ for a sequence of values of $\tau$ progressively tuning the system towards the QCP in the case when $\Lambda_F = \Lambda$. The value of $z$ can be read off by fitting the power law (see the main text). The value $z\approx 2$ following from $B_<$ is clearly visible for nonzero values of the boson fermion coupling $g$ ( $\mathcal{N}_> v_F \propto g^2$). } 
\label{l0_0_z_plots}
\end{figure} 

Before discussing the scale dependencies arising in the RG flow, we inspect the $T=0$ phase diagram - see Fig.~\ref{phase_diag}. We observe hardly any difference between the situations corresponding to $\Lambda_0=0$ and $\Lambda_0>0$ that would be visible in the scaling of the order parameter as a function of the non-thermal control parameter $\tau$. A similar situation may be anticipated for $T>0$ in the behavior of the critical temperature $T_c(\tau)$. This is because (at least in the Hertz-Millis framework) one has $T_c\sim \tau^\psi$ with $\psi=z/(d+z-2)$, which for $d=2$ yields $\psi=1$ irrespective of the value of $z$. As far as the behavior of $T_c$ is concerned,  the distinction between $z=2$ and $z=3$ is only revealed at the level of logarithmic corrections\cite{Millis_1993}. In a realistic (experimental or simulation) situation the value of $z$ may, for example, be read off from the scaling behavior of the correlation function above the quantum critical point (see e.g \cite{Schattner_2016}).  

While presence of the term $B_<$ has no impact on the phase diagram (see Fig.~\ref{phase_diag}), switching on $g$ leads to scaling of the order parameter characterized by $z\approx 3$ in absence of $B_<$ and $z\approx 2$ when $B_<$ is present. At vanishingly low cutoff scale, deviation from the $z=2$ scaling towards higher values of $z$ is clearly visible in the right plot in Fig.~\ref{l0_0_z_plots}. Despite high numerical accuracy, for the considered parameter values, we were however not able to obtain a scaling regime, which could be trustfully interpreted as capturing the behavior corresponding to $z=3$. This emergent picture seems to be consistent with, and offer a potential explanation to results of quantum Monte Carlo simulations of certain models of the electronic nematic\cite{Schattner_2016} as well as the itinerant ferromagnetic\cite{Liu_2022} quantum critical points  which provide evidence for $z=2$-type scaling behavior, rather than  the conventionally anticipated behavior corresponding to $z=3$. 

\emph{Conclusion and perspective}
Within the nonperturbative RG framework, we have derived a generalization of the Hertz action pertinent to fermionic quantum critical systems in $d=2$ characterized by the ordering wavevector $\vec{Q}= \vec{0}$. In our approach both the fermionic and bosonic degrees of freedom are regularized and integrated out of the partition function in parallel, which uncovers a new term in the corresponding order-parameter action and gives rise to a broad scaling regime characterized by the dynamical exponent $z\approx 2$. Our results indicate that a consistent formulation of the momentum-shell Wilsonian RG approach to this problem necessarily requires treating the ordering wavevector as a flowing quantity, which scales to zero exclusively in the infrared limit.  
  The present study addresses only  properties pertinent to the order parameter degrees of freedom and the structure of the bosonic effective action. An  extension accounting for the feedback of bosonic fluctuations on fermionic properties (i.e. the flowing fermion self-energy) can be naturally achieved within our framework, but requires a fully numerical treatment and self-consistent evaluation of the interplay of the two flowing propagators.

\begin{acknowledgments}
 We are grateful to Nicolas Dupuis and Walter Metzner for very useful discussions as well as remarks on the manuscript. M.H. and P.J. acknowledge support from the Polish National Science Center via grants 2021/43/B/ST3/01223 and 2017/26/E/ST3/00211. H.Y. was supported by JSPS KAKENHI Grant No.JP20H01856 and
World Premier International  Research Center Initiative (WPI), MEXT, Japan.

\end{acknowledgments}

\bibliography{bibliography.bib}

\begin{thebibliography}{42}%
\makeatletter
\providecommand \@ifxundefined [1]{%
 \@ifx{#1\undefined}
}%
\providecommand \@ifnum [1]{%
 \ifnum #1\expandafter \@firstoftwo
 \else \expandafter \@secondoftwo
 \fi
}%
\providecommand \@ifx [1]{%
 \ifx #1\expandafter \@firstoftwo
 \else \expandafter \@secondoftwo
 \fi
}%
\providecommand \natexlab [1]{#1}%
\providecommand \enquote  [1]{``#1''}%
\providecommand \bibnamefont  [1]{#1}%
\providecommand \bibfnamefont [1]{#1}%
\providecommand \citenamefont [1]{#1}%
\providecommand \href@noop [0]{\@secondoftwo}%
\providecommand \href [0]{\begingroup \@sanitize@url \@href}%
\providecommand \@href[1]{\@@startlink{#1}\@@href}%
\providecommand \@@href[1]{\endgroup#1\@@endlink}%
\providecommand \@sanitize@url [0]{\catcode `\\12\catcode `\$12\catcode `\&12\catcode `\#12\catcode `\^12\catcode `\_12\catcode `\%12\relax}%
\providecommand \@@startlink[1]{}%
\providecommand \@@endlink[0]{}%
\providecommand \url  [0]{\begingroup\@sanitize@url \@url }%
\providecommand \@url [1]{\endgroup\@href {#1}{\urlprefix }}%
\providecommand \urlprefix  [0]{URL }%
\providecommand \Eprint [0]{\href }%
\providecommand \doibase [0]{http://dx.doi.org/}%
\providecommand \selectlanguage [0]{\@gobble}%
\providecommand \bibinfo  [0]{\@secondoftwo}%
\providecommand \bibfield  [0]{\@secondoftwo}%
\providecommand \translation [1]{[#1]}%
\providecommand \BibitemOpen [0]{}%
\providecommand \bibitemStop [0]{}%
\providecommand \bibitemNoStop [0]{.\EOS\space}%
\providecommand \EOS [0]{\spacefactor3000\relax}%
\providecommand \BibitemShut  [1]{\csname bibitem#1\endcsname}%
\let\auto@bib@innerbib\@empty
\bibitem [{\citenamefont {Keimer}\ \emph {et~al.}(2015)\citenamefont {Keimer}, \citenamefont {Kivelson}, \citenamefont {Norman}, \citenamefont {Uchida},\ and\ \citenamefont {Zaanen}}]{Keimer_2015}%
  \BibitemOpen
  \bibfield  {author} {\bibinfo {author} {\bibfnamefont {B.}~\bibnamefont {Keimer}}, \bibinfo {author} {\bibfnamefont {S.~A.}\ \bibnamefont {Kivelson}}, \bibinfo {author} {\bibfnamefont {M.~R.}\ \bibnamefont {Norman}}, \bibinfo {author} {\bibfnamefont {S.}~\bibnamefont {Uchida}}, \ and\ \bibinfo {author} {\bibfnamefont {J.}~\bibnamefont {Zaanen}},\ }\href@noop {} {\bibfield  {journal} {\bibinfo  {journal} {Nature}\ }\textbf {\bibinfo {volume} {518}},\ \bibinfo {pages} {179} (\bibinfo {year} {2015})}\BibitemShut {NoStop}%
\bibitem [{\citenamefont {Hertz}(1976)}]{Hertz_1976}%
  \BibitemOpen
  \bibfield  {author} {\bibinfo {author} {\bibfnamefont {J.~A.}\ \bibnamefont {Hertz}},\ }\href {\doibase 10.1103/PhysRevB.14.1165} {\bibfield  {journal} {\bibinfo  {journal} {Phys. Rev. B}\ }\textbf {\bibinfo {volume} {14}},\ \bibinfo {pages} {1165} (\bibinfo {year} {1976})}\BibitemShut {NoStop}%
\bibitem [{\citenamefont {Millis}(1993)}]{Millis_1993}%
  \BibitemOpen
  \bibfield  {author} {\bibinfo {author} {\bibfnamefont {A.~J.}\ \bibnamefont {Millis}},\ }\href {\doibase 10.1103/PhysRevB.48.7183} {\bibfield  {journal} {\bibinfo  {journal} {Phys. Rev. B}\ }\textbf {\bibinfo {volume} {48}},\ \bibinfo {pages} {7183} (\bibinfo {year} {1993})}\BibitemShut {NoStop}%
\bibitem [{\citenamefont {Cardy}(1996)}]{Cardy_1996}%
  \BibitemOpen
  \bibfield  {author} {\bibinfo {author} {\bibfnamefont {J.}~\bibnamefont {Cardy}},\ }\href {\doibase 10.1017/CBO9781316036440} {\emph {\bibinfo {title} {{Scaling and Renormalization in Statistical Physics}}}}\ (\bibinfo  {publisher} {Cambridge University Press},\ \bibinfo {year} {1996})\BibitemShut {NoStop}%
\bibitem [{\citenamefont {Goldenfeld}(1992)}]{Goldenfeld_1992}%
  \BibitemOpen
  \bibfield  {author} {\bibinfo {author} {\bibfnamefont {N.}~\bibnamefont {Goldenfeld}},\ }\href {\doibase 10.1201/9780429493492} {\emph {\bibinfo {title} {Lect. Phase Transitions Renorm. Gr.}}},\ \bibinfo {edition} {1st}\ ed.\ (\bibinfo  {publisher} {CRC Press},\ \bibinfo {address} {Boca Raton},\ \bibinfo {year} {1992})\BibitemShut {NoStop}%
\bibitem [{\citenamefont {Nagaosa}(1998)}]{Nagaosa_1998}%
  \BibitemOpen
  \bibfield  {author} {\bibinfo {author} {\bibfnamefont {N.}~\bibnamefont {Nagaosa}},\ }\href {https://www.taylorfrancis.com/books/9780429962042} {\emph {\bibinfo {title} {Quantum Field Theory in Strongly Correlated Electronic Systems}}}\ (\bibinfo  {publisher} {Springer-Verlag},\ \bibinfo {year} {1998})\BibitemShut {NoStop}%
\bibitem [{\citenamefont {Belitz}\ \emph {et~al.}(2005)\citenamefont {Belitz}, \citenamefont {Kirkpatrick},\ and\ \citenamefont {Vojta}}]{Belitz_2005}%
  \BibitemOpen
  \bibfield  {author} {\bibinfo {author} {\bibfnamefont {D.}~\bibnamefont {Belitz}}, \bibinfo {author} {\bibfnamefont {T.~R.}\ \bibnamefont {Kirkpatrick}}, \ and\ \bibinfo {author} {\bibfnamefont {T.}~\bibnamefont {Vojta}},\ }\href {\doibase 10.1103/RevModPhys.77.579} {\bibfield  {journal} {\bibinfo  {journal} {Rev. Mod. Phys.}\ }\textbf {\bibinfo {volume} {77}},\ \bibinfo {pages} {579} (\bibinfo {year} {2005})}\BibitemShut {NoStop}%
\bibitem [{\citenamefont {L\"ohneysen}\ \emph {et~al.}(2007)\citenamefont {L\"ohneysen}, \citenamefont {Rosch}, \citenamefont {Vojta},\ and\ \citenamefont {W\"olfle}}]{Lohneysen_2007}%
  \BibitemOpen
  \bibfield  {author} {\bibinfo {author} {\bibfnamefont {H.~v.}\ \bibnamefont {L\"ohneysen}}, \bibinfo {author} {\bibfnamefont {A.}~\bibnamefont {Rosch}}, \bibinfo {author} {\bibfnamefont {M.}~\bibnamefont {Vojta}}, \ and\ \bibinfo {author} {\bibfnamefont {P.}~\bibnamefont {W\"olfle}},\ }\href {\doibase 10.1103/RevModPhys.79.1015} {\bibfield  {journal} {\bibinfo  {journal} {Rev. Mod. Phys.}\ }\textbf {\bibinfo {volume} {79}},\ \bibinfo {pages} {1015} (\bibinfo {year} {2007})}\BibitemShut {NoStop}%
\bibitem [{\citenamefont {Metlitski}\ and\ \citenamefont {Sachdev}(2010)}]{Metlitski_1010}%
  \BibitemOpen
  \bibfield  {author} {\bibinfo {author} {\bibfnamefont {M.~A.}\ \bibnamefont {Metlitski}}\ and\ \bibinfo {author} {\bibfnamefont {S.}~\bibnamefont {Sachdev}},\ }\href {\doibase 10.1103/PhysRevB.82.075128} {\bibfield  {journal} {\bibinfo  {journal} {Phys. Rev. B}\ }\textbf {\bibinfo {volume} {82}},\ \bibinfo {pages} {075128} (\bibinfo {year} {2010})}\BibitemShut {NoStop}%
\bibitem [{\citenamefont {Thier}\ and\ \citenamefont {Metzner}(2011)}]{Thier_2011}%
  \BibitemOpen
  \bibfield  {author} {\bibinfo {author} {\bibfnamefont {S.~C.}\ \bibnamefont {Thier}}\ and\ \bibinfo {author} {\bibfnamefont {W.}~\bibnamefont {Metzner}},\ }\href {\doibase 10.1103/PhysRevB.84.155133} {\bibfield  {journal} {\bibinfo  {journal} {Phys. Rev. B}\ }\textbf {\bibinfo {volume} {84}},\ \bibinfo {pages} {155133} (\bibinfo {year} {2011})}\BibitemShut {NoStop}%
\bibitem [{\citenamefont {Lee}(2009)}]{Lee_2009}%
  \BibitemOpen
  \bibfield  {author} {\bibinfo {author} {\bibfnamefont {S.-S.}\ \bibnamefont {Lee}},\ }\href {\doibase 10.1103/PhysRevB.80.165102} {\bibfield  {journal} {\bibinfo  {journal} {Phys. Rev. B}\ }\textbf {\bibinfo {volume} {80}},\ \bibinfo {pages} {165102} (\bibinfo {year} {2009})}\BibitemShut {NoStop}%
\bibitem [{\citenamefont {Maslov}\ and\ \citenamefont {Chubukov}(2010)}]{Maslow_2010}%
  \BibitemOpen
  \bibfield  {author} {\bibinfo {author} {\bibfnamefont {D.~L.}\ \bibnamefont {Maslov}}\ and\ \bibinfo {author} {\bibfnamefont {A.~V.}\ \bibnamefont {Chubukov}},\ }\href {\doibase 10.1103/PhysRevB.81.045110} {\bibfield  {journal} {\bibinfo  {journal} {Phys. Rev. B}\ }\textbf {\bibinfo {volume} {81}},\ \bibinfo {pages} {045110} (\bibinfo {year} {2010})}\BibitemShut {NoStop}%
\bibitem [{\citenamefont {Mross}\ \emph {et~al.}(2010)\citenamefont {Mross}, \citenamefont {McGreevy}, \citenamefont {Liu},\ and\ \citenamefont {Senthil}}]{Mross_2010}%
  \BibitemOpen
  \bibfield  {author} {\bibinfo {author} {\bibfnamefont {D.~F.}\ \bibnamefont {Mross}}, \bibinfo {author} {\bibfnamefont {J.}~\bibnamefont {McGreevy}}, \bibinfo {author} {\bibfnamefont {H.}~\bibnamefont {Liu}}, \ and\ \bibinfo {author} {\bibfnamefont {T.}~\bibnamefont {Senthil}},\ }\href {\doibase 10.1103/PhysRevB.82.045121} {\bibfield  {journal} {\bibinfo  {journal} {Phys. Rev. B}\ }\textbf {\bibinfo {volume} {82}},\ \bibinfo {pages} {045121} (\bibinfo {year} {2010})}\BibitemShut {NoStop}%
\bibitem [{\citenamefont {Drukier}\ \emph {et~al.}(2012)\citenamefont {Drukier}, \citenamefont {Bartosch}, \citenamefont {Isidori},\ and\ \citenamefont {Kopietz}}]{Drukier_2012}%
  \BibitemOpen
  \bibfield  {author} {\bibinfo {author} {\bibfnamefont {C.}~\bibnamefont {Drukier}}, \bibinfo {author} {\bibfnamefont {L.}~\bibnamefont {Bartosch}}, \bibinfo {author} {\bibfnamefont {A.}~\bibnamefont {Isidori}}, \ and\ \bibinfo {author} {\bibfnamefont {P.}~\bibnamefont {Kopietz}},\ }\href {\doibase 10.1103/PhysRevB.85.245120} {\bibfield  {journal} {\bibinfo  {journal} {Phys. Rev. B}\ }\textbf {\bibinfo {volume} {85}},\ \bibinfo {pages} {245120} (\bibinfo {year} {2012})}\BibitemShut {NoStop}%
\bibitem [{\citenamefont {Fitzpatrick}\ \emph {et~al.}(2013)\citenamefont {Fitzpatrick}, \citenamefont {Kachru}, \citenamefont {Kaplan},\ and\ \citenamefont {Raghu}}]{Fitzpatrick_2013}%
  \BibitemOpen
  \bibfield  {author} {\bibinfo {author} {\bibfnamefont {A.~L.}\ \bibnamefont {Fitzpatrick}}, \bibinfo {author} {\bibfnamefont {S.}~\bibnamefont {Kachru}}, \bibinfo {author} {\bibfnamefont {J.}~\bibnamefont {Kaplan}}, \ and\ \bibinfo {author} {\bibfnamefont {S.}~\bibnamefont {Raghu}},\ }\href {\doibase 10.1103/PhysRevB.88.125116} {\bibfield  {journal} {\bibinfo  {journal} {Phys. Rev. B}\ }\textbf {\bibinfo {volume} {88}},\ \bibinfo {pages} {125116} (\bibinfo {year} {2013})}\BibitemShut {NoStop}%
\bibitem [{\citenamefont {Dalidovich}\ and\ \citenamefont {Lee}(2013)}]{Dalidovich_2013}%
  \BibitemOpen
  \bibfield  {author} {\bibinfo {author} {\bibfnamefont {D.}~\bibnamefont {Dalidovich}}\ and\ \bibinfo {author} {\bibfnamefont {S.-S.}\ \bibnamefont {Lee}},\ }\href {\doibase 10.1103/PhysRevB.88.245106} {\bibfield  {journal} {\bibinfo  {journal} {Phys. Rev. B}\ }\textbf {\bibinfo {volume} {88}},\ \bibinfo {pages} {245106} (\bibinfo {year} {2013})}\BibitemShut {NoStop}%
\bibitem [{\citenamefont {Mandal}\ and\ \citenamefont {Lee}(2015)}]{Mandal_2015}%
  \BibitemOpen
  \bibfield  {author} {\bibinfo {author} {\bibfnamefont {I.}~\bibnamefont {Mandal}}\ and\ \bibinfo {author} {\bibfnamefont {S.-S.}\ \bibnamefont {Lee}},\ }\href {\doibase 10.1103/PhysRevB.92.035141} {\bibfield  {journal} {\bibinfo  {journal} {Phys. Rev. B}\ }\textbf {\bibinfo {volume} {92}},\ \bibinfo {pages} {035141} (\bibinfo {year} {2015})}\BibitemShut {NoStop}%
\bibitem [{\citenamefont {Holder}\ and\ \citenamefont {Metzner}(2015)}]{Holder_2015}%
  \BibitemOpen
  \bibfield  {author} {\bibinfo {author} {\bibfnamefont {T.}~\bibnamefont {Holder}}\ and\ \bibinfo {author} {\bibfnamefont {W.}~\bibnamefont {Metzner}},\ }\href {\doibase 10.1103/PhysRevB.92.041112} {\bibfield  {journal} {\bibinfo  {journal} {Phys. Rev. B}\ }\textbf {\bibinfo {volume} {92}},\ \bibinfo {pages} {041112} (\bibinfo {year} {2015})}\BibitemShut {NoStop}%
\bibitem [{\citenamefont {Ridgway}\ and\ \citenamefont {Hooley}(2015)}]{Ridgway_2015}%
  \BibitemOpen
  \bibfield  {author} {\bibinfo {author} {\bibfnamefont {S.~P.}\ \bibnamefont {Ridgway}}\ and\ \bibinfo {author} {\bibfnamefont {C.~A.}\ \bibnamefont {Hooley}},\ }\href {\doibase 10.1103/PhysRevLett.114.226404} {\bibfield  {journal} {\bibinfo  {journal} {Phys. Rev. Lett.}\ }\textbf {\bibinfo {volume} {114}},\ \bibinfo {pages} {226404} (\bibinfo {year} {2015})}\BibitemShut {NoStop}%
\bibitem [{\citenamefont {Punk}(2016)}]{Punk_2016}%
  \BibitemOpen
  \bibfield  {author} {\bibinfo {author} {\bibfnamefont {M.}~\bibnamefont {Punk}},\ }\href {\doibase 10.1103/PhysRevB.94.195113} {\bibfield  {journal} {\bibinfo  {journal} {Phys. Rev. B}\ }\textbf {\bibinfo {volume} {94}},\ \bibinfo {pages} {195113} (\bibinfo {year} {2016})}\BibitemShut {NoStop}%
\bibitem [{\citenamefont {Trott}\ and\ \citenamefont {Hooley}(2018)}]{Trott_2018}%
  \BibitemOpen
  \bibfield  {author} {\bibinfo {author} {\bibfnamefont {M.~J.}\ \bibnamefont {Trott}}\ and\ \bibinfo {author} {\bibfnamefont {C.~A.}\ \bibnamefont {Hooley}},\ }\href {\doibase 10.1103/PhysRevB.98.201113} {\bibfield  {journal} {\bibinfo  {journal} {Phys. Rev. B}\ }\textbf {\bibinfo {volume} {98}},\ \bibinfo {pages} {201113} (\bibinfo {year} {2018})}\BibitemShut {NoStop}%
\bibitem [{\citenamefont {Lee}(2018)}]{Lee_2018}%
  \BibitemOpen
  \bibfield  {author} {\bibinfo {author} {\bibfnamefont {S.-S.}\ \bibnamefont {Lee}},\ }\href {\doibase https://doi.org/10.1146/annurev-conmatphys-031016-025531} {\bibfield  {journal} {\bibinfo  {journal} {Annual Review of Condensed Matter Physics}\ }\textbf {\bibinfo {volume} {9}},\ \bibinfo {pages} {227} (\bibinfo {year} {2018})}\BibitemShut {NoStop}%
\bibitem [{\citenamefont {Damia}\ \emph {et~al.}(2019)\citenamefont {Damia}, \citenamefont {Kachru}, \citenamefont {Raghu},\ and\ \citenamefont {Torroba}}]{Damia_2019}%
  \BibitemOpen
  \bibfield  {author} {\bibinfo {author} {\bibfnamefont {J.~A.}\ \bibnamefont {Damia}}, \bibinfo {author} {\bibfnamefont {S.}~\bibnamefont {Kachru}}, \bibinfo {author} {\bibfnamefont {S.}~\bibnamefont {Raghu}}, \ and\ \bibinfo {author} {\bibfnamefont {G.}~\bibnamefont {Torroba}},\ }\href {\doibase 10.1103/PhysRevLett.123.096402} {\bibfield  {journal} {\bibinfo  {journal} {Phys. Rev. Lett.}\ }\textbf {\bibinfo {volume} {123}},\ \bibinfo {pages} {096402} (\bibinfo {year} {2019})}\BibitemShut {NoStop}%
\bibitem [{\citenamefont {Damia}\ \emph {et~al.}(2020)\citenamefont {Damia}, \citenamefont {Sol\'{\i}s},\ and\ \citenamefont {Torroba}}]{Damia_2020}%
  \BibitemOpen
  \bibfield  {author} {\bibinfo {author} {\bibfnamefont {J.~A.}\ \bibnamefont {Damia}}, \bibinfo {author} {\bibfnamefont {M.}~\bibnamefont {Sol\'{\i}s}}, \ and\ \bibinfo {author} {\bibfnamefont {G.}~\bibnamefont {Torroba}},\ }\href {\doibase 10.1103/PhysRevB.102.045147} {\bibfield  {journal} {\bibinfo  {journal} {Phys. Rev. B}\ }\textbf {\bibinfo {volume} {102}},\ \bibinfo {pages} {045147} (\bibinfo {year} {2020})}\BibitemShut {NoStop}%
\bibitem [{\citenamefont {Säterskog}(2021)}]{Saterskog_2021}%
  \BibitemOpen
  \bibfield  {author} {\bibinfo {author} {\bibfnamefont {P.}~\bibnamefont {Säterskog}},\ }\href {\doibase 10.21468/SciPostPhys.10.3.067} {\bibfield  {journal} {\bibinfo  {journal} {SciPost Phys.}\ }\textbf {\bibinfo {volume} {10}},\ \bibinfo {pages} {067} (\bibinfo {year} {2021})}\BibitemShut {NoStop}%
\bibitem [{\citenamefont {Zhang}\ and\ \citenamefont {Chen}(2023)}]{Zhang_2023}%
  \BibitemOpen
  \bibfield  {author} {\bibinfo {author} {\bibfnamefont {X.-T.}\ \bibnamefont {Zhang}}\ and\ \bibinfo {author} {\bibfnamefont {G.}~\bibnamefont {Chen}},\ }\href@noop {} {\bibfield  {journal} {\bibinfo  {journal} {npj Quantum Materials}\ }\textbf {\bibinfo {volume} {8}},\ \bibinfo {pages} {10} (\bibinfo {year} {2023})}\BibitemShut {NoStop}%
\bibitem [{\citenamefont {Mayrhofer}\ \emph {et~al.}(2024)\citenamefont {Mayrhofer}, \citenamefont {Wölfle},\ and\ \citenamefont {Chubukov}}]{Mayrhofer_2024}%
  \BibitemOpen
  \bibfield  {author} {\bibinfo {author} {\bibfnamefont {R.~D.}\ \bibnamefont {Mayrhofer}}, \bibinfo {author} {\bibfnamefont {P.}~\bibnamefont {Wölfle}}, \ and\ \bibinfo {author} {\bibfnamefont {A.~V.}\ \bibnamefont {Chubukov}},\ }\href@noop {} {} (\bibinfo {year} {2024}),\ \Eprint {http://arxiv.org/abs/2403.09835} {arXiv:2403.09835 [cond-mat.str-el]} \BibitemShut {NoStop}%
\bibitem [{\citenamefont {Wetterich}(1993)}]{Wetterich_1993}%
  \BibitemOpen
  \bibfield  {author} {\bibinfo {author} {\bibfnamefont {C.}~\bibnamefont {Wetterich}},\ }\href {\doibase 10.1016/0370-2693(93)90726-X} {\bibfield  {journal} {\bibinfo  {journal} {Phys. Lett. B}\ }\textbf {\bibinfo {volume} {301}},\ \bibinfo {pages} {90} (\bibinfo {year} {1993})}\BibitemShut {NoStop}%
\bibitem [{\citenamefont {Kopietz}\ \emph {et~al.}(2010)\citenamefont {Kopietz}, \citenamefont {Bartosch},\ and\ \citenamefont {Sch{\"{u}}tz}}]{Kopietz_book}%
  \BibitemOpen
  \bibfield  {author} {\bibinfo {author} {\bibfnamefont {P.}~\bibnamefont {Kopietz}}, \bibinfo {author} {\bibfnamefont {L.}~\bibnamefont {Bartosch}}, \ and\ \bibinfo {author} {\bibfnamefont {F.}~\bibnamefont {Sch{\"{u}}tz}},\ }\href {\doibase doi.org/10.1007/978-3-642-05094-7} {\emph {\bibinfo {title} {{Introduction to the Functional Renormalization Group}}}},\ \bibinfo {edition} {1st}\ ed.\ (\bibinfo  {publisher} {Springer Berlin},\ \bibinfo {address} {Heidelberg},\ \bibinfo {year} {2010})\BibitemShut {NoStop}%
\bibitem [{\citenamefont {Dupuis}\ \emph {et~al.}(2021)\citenamefont {Dupuis}, \citenamefont {Canet}, \citenamefont {Eichhorn}, \citenamefont {Metzner}, \citenamefont {Pawlowski}, \citenamefont {Tissier},\ and\ \citenamefont {Wschebor}}]{Dupuis_2021}%
  \BibitemOpen
  \bibfield  {author} {\bibinfo {author} {\bibfnamefont {N.}~\bibnamefont {Dupuis}}, \bibinfo {author} {\bibfnamefont {L.}~\bibnamefont {Canet}}, \bibinfo {author} {\bibfnamefont {A.}~\bibnamefont {Eichhorn}}, \bibinfo {author} {\bibfnamefont {W.}~\bibnamefont {Metzner}}, \bibinfo {author} {\bibfnamefont {J.}~\bibnamefont {Pawlowski}}, \bibinfo {author} {\bibfnamefont {M.}~\bibnamefont {Tissier}}, \ and\ \bibinfo {author} {\bibfnamefont {N.}~\bibnamefont {Wschebor}},\ }\href {\doibase https://doi.org/10.1016/j.physrep.2021.01.001} {\bibfield  {journal} {\bibinfo  {journal} {Physics Reports}\ }\textbf {\bibinfo {volume} {910}},\ \bibinfo {pages} {1} (\bibinfo {year} {2021})}\BibitemShut {NoStop}%
\bibitem [{\citenamefont {Canet}\ \emph {et~al.}(2010)\citenamefont {Canet}, \citenamefont {Chat\'e}, \citenamefont {Delamotte},\ and\ \citenamefont {Wschebor}}]{Canet_2010}%
  \BibitemOpen
  \bibfield  {author} {\bibinfo {author} {\bibfnamefont {L.}~\bibnamefont {Canet}}, \bibinfo {author} {\bibfnamefont {H.}~\bibnamefont {Chat\'e}}, \bibinfo {author} {\bibfnamefont {B.}~\bibnamefont {Delamotte}}, \ and\ \bibinfo {author} {\bibfnamefont {N.}~\bibnamefont {Wschebor}},\ }\href {\doibase 10.1103/PhysRevLett.104.150601} {\bibfield  {journal} {\bibinfo  {journal} {Phys. Rev. Lett.}\ }\textbf {\bibinfo {volume} {104}},\ \bibinfo {pages} {150601} (\bibinfo {year} {2010})}\BibitemShut {NoStop}%
\bibitem [{\citenamefont {Tissier}\ and\ \citenamefont {Tarjus}(2011)}]{Tissier_2011}%
  \BibitemOpen
  \bibfield  {author} {\bibinfo {author} {\bibfnamefont {M.}~\bibnamefont {Tissier}}\ and\ \bibinfo {author} {\bibfnamefont {G.}~\bibnamefont {Tarjus}},\ }\href {\doibase 10.1103/PhysRevLett.107.041601} {\bibfield  {journal} {\bibinfo  {journal} {Phys. Rev. Lett.}\ }\textbf {\bibinfo {volume} {107}},\ \bibinfo {pages} {041601} (\bibinfo {year} {2011})}\BibitemShut {NoStop}%
\bibitem [{\citenamefont {Yabunaka}\ and\ \citenamefont {Delamotte}(2017)}]{Yabunaka_2017}%
  \BibitemOpen
  \bibfield  {author} {\bibinfo {author} {\bibfnamefont {S.}~\bibnamefont {Yabunaka}}\ and\ \bibinfo {author} {\bibfnamefont {B.}~\bibnamefont {Delamotte}},\ }\href {\doibase 10.1103/PhysRevLett.119.191602} {\bibfield  {journal} {\bibinfo  {journal} {Phys. Rev. Lett.}\ }\textbf {\bibinfo {volume} {119}},\ \bibinfo {pages} {191602} (\bibinfo {year} {2017})}\BibitemShut {NoStop}%
\bibitem [{\citenamefont {Chlebicki}\ and\ \citenamefont {Jakubczyk}(2021)}]{Chlebicki_2021}%
  \BibitemOpen
  \bibfield  {author} {\bibinfo {author} {\bibfnamefont {A.}~\bibnamefont {Chlebicki}}\ and\ \bibinfo {author} {\bibfnamefont {P.}~\bibnamefont {Jakubczyk}},\ }\href {\doibase 10.21468/SciPostPhys.10.6.134} {\bibfield  {journal} {\bibinfo  {journal} {SciPost Phys.}\ }\textbf {\bibinfo {volume} {10}},\ \bibinfo {pages} {134} (\bibinfo {year} {2021})}\BibitemShut {NoStop}%
\bibitem [{\citenamefont {Obert}\ \emph {et~al.}(2013)\citenamefont {Obert}, \citenamefont {Husemann},\ and\ \citenamefont {Metzner}}]{Obert_2013}%
  \BibitemOpen
  \bibfield  {author} {\bibinfo {author} {\bibfnamefont {B.}~\bibnamefont {Obert}}, \bibinfo {author} {\bibfnamefont {C.}~\bibnamefont {Husemann}}, \ and\ \bibinfo {author} {\bibfnamefont {W.}~\bibnamefont {Metzner}},\ }\href {\doibase 10.1103/PhysRevB.88.144508} {\bibfield  {journal} {\bibinfo  {journal} {Phys. Rev. B}\ }\textbf {\bibinfo {volume} {88}},\ \bibinfo {pages} {144508} (\bibinfo {year} {2013})}\BibitemShut {NoStop}%
\bibitem [{\citenamefont {Obert}(2014)}]{Obert_2014_PhD}%
  \BibitemOpen
  \bibfield  {author} {\bibinfo {author} {\bibfnamefont {B.}~\bibnamefont {Obert}},\ }\href {https://elib.uni-stuttgart.de/bitstream/11682/6862/1/Doktorarbeit_Bibliothek.pdf} {\emph {\bibinfo {title} {{Renormalization group analysis of order parameter fluctuations in fermionic superfluids}}}}\ (\bibinfo  {publisher} {Ph.D Thesis, University of Stuttgart},\ \bibinfo {year} {2014})\BibitemShut {NoStop}%
\bibitem [{\citenamefont {Berges}\ \emph {et~al.}(2002)\citenamefont {Berges}, \citenamefont {Tetradis},\ and\ \citenamefont {Wetterich}}]{Berges_2002}%
  \BibitemOpen
  \bibfield  {author} {\bibinfo {author} {\bibfnamefont {J.}~\bibnamefont {Berges}}, \bibinfo {author} {\bibfnamefont {N.}~\bibnamefont {Tetradis}}, \ and\ \bibinfo {author} {\bibfnamefont {C.}~\bibnamefont {Wetterich}},\ }\href {\doibase 10.1016/S0370-1573(01)00098-9} {\bibfield  {journal} {\bibinfo  {journal} {Phys. Rep.}\ }\textbf {\bibinfo {volume} {363}},\ \bibinfo {pages} {223} (\bibinfo {year} {2002})},\ \Eprint {http://arxiv.org/abs/0005122} {0005122 [hep-ph]} \BibitemShut {NoStop}%
\bibitem [{\citenamefont {Jakubczyk}\ \emph {et~al.}(2008)\citenamefont {Jakubczyk}, \citenamefont {Strack}, \citenamefont {Katanin},\ and\ \citenamefont {Metzner}}]{Jakubczyk_2008}%
  \BibitemOpen
  \bibfield  {author} {\bibinfo {author} {\bibfnamefont {P.}~\bibnamefont {Jakubczyk}}, \bibinfo {author} {\bibfnamefont {P.}~\bibnamefont {Strack}}, \bibinfo {author} {\bibfnamefont {A.~A.}\ \bibnamefont {Katanin}}, \ and\ \bibinfo {author} {\bibfnamefont {W.}~\bibnamefont {Metzner}},\ }\href {\doibase 10.1103/PhysRevB.77.195120} {\bibfield  {journal} {\bibinfo  {journal} {Phys. Rev. B}\ }\textbf {\bibinfo {volume} {77}},\ \bibinfo {pages} {195120} (\bibinfo {year} {2008})}\BibitemShut {NoStop}%
\bibitem [{\citenamefont {Bauer}\ \emph {et~al.}(2011)\citenamefont {Bauer}, \citenamefont {Jakubczyk},\ and\ \citenamefont {Metzner}}]{Bauer_2011}%
  \BibitemOpen
  \bibfield  {author} {\bibinfo {author} {\bibfnamefont {J.}~\bibnamefont {Bauer}}, \bibinfo {author} {\bibfnamefont {P.}~\bibnamefont {Jakubczyk}}, \ and\ \bibinfo {author} {\bibfnamefont {W.}~\bibnamefont {Metzner}},\ }\href {\doibase 10.1103/PhysRevB.84.075122} {\bibfield  {journal} {\bibinfo  {journal} {Phys. Rev. B}\ }\textbf {\bibinfo {volume} {84}},\ \bibinfo {pages} {075122} (\bibinfo {year} {2011})}\BibitemShut {NoStop}%
\bibitem [{\citenamefont {Litim}(2001)}]{Litim_2001}%
  \BibitemOpen
  \bibfield  {author} {\bibinfo {author} {\bibfnamefont {D.~F.}\ \bibnamefont {Litim}},\ }\href {\doibase 10.1103/PhysRevD.64.105007} {\bibfield  {journal} {\bibinfo  {journal} {Phys. Rev. D}\ }\textbf {\bibinfo {volume} {64}},\ \bibinfo {pages} {105007} (\bibinfo {year} {2001})}\BibitemShut {NoStop}%
\bibitem [{\citenamefont {Schattner}\ \emph {et~al.}(2016)\citenamefont {Schattner}, \citenamefont {Lederer}, \citenamefont {Kivelson},\ and\ \citenamefont {Berg}}]{Schattner_2016}%
  \BibitemOpen
  \bibfield  {author} {\bibinfo {author} {\bibfnamefont {Y.}~\bibnamefont {Schattner}}, \bibinfo {author} {\bibfnamefont {S.}~\bibnamefont {Lederer}}, \bibinfo {author} {\bibfnamefont {S.~A.}\ \bibnamefont {Kivelson}}, \ and\ \bibinfo {author} {\bibfnamefont {E.}~\bibnamefont {Berg}},\ }\href {\doibase 10.1103/PhysRevX.6.031028} {\bibfield  {journal} {\bibinfo  {journal} {Phys. Rev. X}\ }\textbf {\bibinfo {volume} {6}},\ \bibinfo {pages} {031028} (\bibinfo {year} {2016})}\BibitemShut {NoStop}%
\bibitem [{\citenamefont {Liu}\ \emph {et~al.}(2022)\citenamefont {Liu}, \citenamefont {Jiang}, \citenamefont {Klein}, \citenamefont {Wang}, \citenamefont {Sun}, \citenamefont {Chubukov},\ and\ \citenamefont {Meng}}]{Liu_2022}%
  \BibitemOpen
  \bibfield  {author} {\bibinfo {author} {\bibfnamefont {Y.}~\bibnamefont {Liu}}, \bibinfo {author} {\bibfnamefont {W.}~\bibnamefont {Jiang}}, \bibinfo {author} {\bibfnamefont {A.}~\bibnamefont {Klein}}, \bibinfo {author} {\bibfnamefont {Y.}~\bibnamefont {Wang}}, \bibinfo {author} {\bibfnamefont {K.}~\bibnamefont {Sun}}, \bibinfo {author} {\bibfnamefont {A.~V.}\ \bibnamefont {Chubukov}}, \ and\ \bibinfo {author} {\bibfnamefont {Z.~Y.}\ \bibnamefont {Meng}},\ }\href {\doibase 10.1103/PhysRevB.105.L041111} {\bibfield  {journal} {\bibinfo  {journal} {Phys. Rev. B}\ }\textbf {\bibinfo {volume} {105}},\ \bibinfo {pages} {L041111} (\bibinfo {year} {2022})}\BibitemShut {NoStop}%
\end{thebibliography}%
\bibliographystyle{apsrev4-1}

\clearpage
\onecolumngrid

\begin{center}
\textbf{\large Supplemental Material}
\end{center}

\section*{Evaluation of $\mathcal{X}(q, \, \Lambda_F)$}

In this section we present the calculation of $\mathcal{X}(q, \, \Lambda_F)$ - see Eq.(10) in the main text. We start with an explicit formula:

\begin{align}
    \mathcal{X}(q, \, \Lambda_F) = 
    - \frac{2 g^2}{(2 \pi)^3} 
    \int_{-\infty}^{\infty} {\rm d}k_0 
    \int {\rm d}^2 k 
    \frac{ \partial_\Lambda R_f(\vec{k}) }{ \big(-ik_0+ f(\vec{k}) \big)^2} 
    \Bigg[
        \frac{1}{-i (k_0 + q_0) + f(\vec{k} + \vec{q})}
        + 
        \frac{1}{-i (k_0 - q_0) + f(\vec{k} - \vec{q})}
    \Bigg] \; ,
\end{align}
\noindent where we introduced the regularized fermionic dispersion $f(\vec{k}) = \xi_{\vec{k}}+R_f(\vec{k})$. The term $\partial_\Lambda R_f$ restricts the region of integration over momenta to the shell $|\vec{k}|\in ( k_F -\Lambda_F, k_F + \Lambda_F )$, such that:

\begin{align}
    \mathcal{X}(q, q_0, \Lambda_F) = 
    \frac{2 \, i \, g^2 v_F}{(2 \pi)^3} 
    \Bigg\{
        \Big(1 + \frac{\Lambda_F}{k_F} \Big)
        \int_{-\infty}^{\infty} {\rm d}k_0
        \int_{k_F}^{k_F + \Lambda_F} {\rm d} k \int_0^{2 \pi} {\rm d}\phi 
        \frac{ k }{ \big( k_0 + i \, \xi_{k_F + \Lambda_F} \big)^2} 
        \Bigg[
            \frac{1}{ k_0 + q_0 + i f(\vec{k} + \vec{q})}
            + 
            \frac{1}{ k_0 - q_0 + i f(\vec{k} + \vec{q})}
        \Bigg] 
        +  \\ \notag
        - \Big( 1 - \frac{\Lambda_F}{k_F} \Big)
        \int_{-\infty}^{\infty} {\rm d}k_0
        \int_{k_F - \Lambda_F}^{k_F } {\rm d} k \int_0^{2 \pi} {\rm d}\phi 
        \frac{ k }{ \big( k_0 +  i \, \xi_{k_F - \Lambda_F} \big)^2} 
        \Bigg[
            \frac{1}{k_0 + q_0 + i f(\vec{k} + \vec{q})}
            + 
            \frac{1}{k_0 - q_0 + i f(\vec{k} + \vec{q})}
        \Bigg]
        \Bigg\} \; .     
\end{align}

\noindent We perform the frequency integration using the residue theorem:

\begin{align}
\label{freq_integrated}
    \mathcal{X}(q, q_0, \Lambda_F) = 
    - \frac{2 g^2 v_F}{(2 \pi)^2}
    \Biggl\{
        \Bigl(1 + \frac{\Lambda_F}{k_F} \Bigr)
        \int_{k_F}^{k_F + \Lambda_F} {\rm d} k 
        \int_0^{2 \pi} {\rm d}\phi \; k \;
        \Biggl[
            \frac{ \Theta \Bigl[ - f(\vec{k} + \vec{q}) \Bigr] }{ \bigl( q_0 + i \, \xi_{k_F + \Lambda_F} - i \, f(\vec{k} + \vec{q}) \bigr)^2}
            + 
            \left( q_0 \to - q_0 \right) \;
        \Biggr] \; 
        +  \\ \notag
        + \, \Bigl(1 - \frac{\Lambda_F}{k_F} \Bigr)
        \int_{k_F - \Lambda_F}^{k_F } {\rm d} k 
        \int_0^{2 \pi} {\rm d}\phi \; k \;
        \Biggl[
            \frac{ \Theta \Bigl[   f(\vec{k} + \vec{q}) \Bigr] }{ \bigl( q_0 + i \, f(\vec{k} + \vec{q}) - i \, \xi_{k_F - \Lambda_F} \bigr)^2}
            + 
            \left( q_0 \to - q_0 \right) \;
        \Biggr]
    \Biggr\} \; .
\end{align}

\noindent The schematic plot illustrating the geometry of the momentum integration is presented in Fig.~\ref{regule}.

\begin{figure}[h]
\includegraphics[width= 0.9\textwidth]{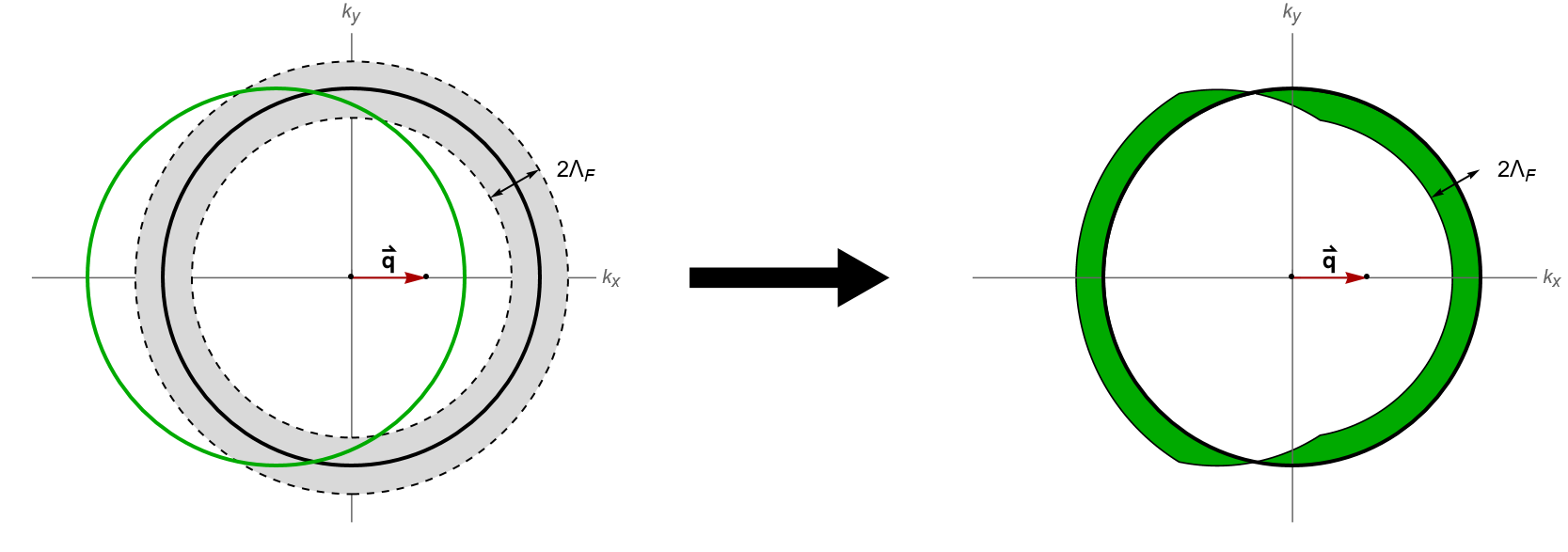}
\caption{Left:The gray shell around the Fermi surface marks the region of momenta integration restricted by the factor $\partial_{\Lambda} R_F$. The Fermi surface shifted by the vector $-\vec{q}$ is represented by green solid line. Right: the ultimate momentum integration region with nonzero contribution (see Eq.(\ref{freq_integrated}). } 
\label{regule}
\end{figure} 

\begin{figure}
\includegraphics[width= \textwidth]{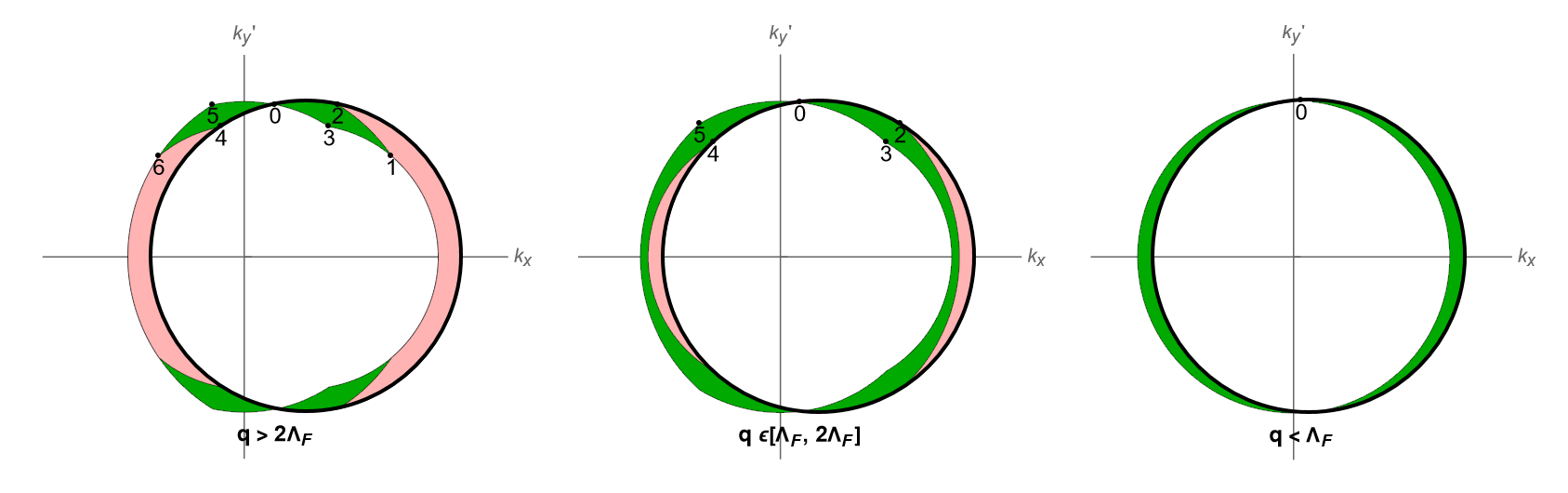}
\caption{ The regions of momentum space where the integrand in Eq. (\ref{freq_integrated}) is nonzero. The integrand is constant for the momentum vector $\vec{k}$ which lies in the green region due to cut-off function which regularizes the fermi dispersion. For a momentum vector which lies in the red region, the cut-off function becomes inactive. Black points define the geometry of the region. } 
\label{geometry}
\end{figure} 


\subsection*{ Momentum space geometry}

The geometry of the integration region depends on the relative ratio $|\vec{q}|/\Lambda_F$, because the regularized fermionic dispersion $f(\vec{k})$ is constant for certain values of momenta. In Fig. \ref{geometry} we present the three possible cases: (a) $q > 2 \Lambda_F$, (b) $q \in ( \Lambda_F , 2 \Lambda_F )$, and (c) $q < \Lambda_F$. From now on $q \coloneqq |\vec{q}|$. Green color denotes the region for which the integrand in Eq.(\ref{freq_integrated}) becomes independent of $\vec{k}$ and red color denotes the region, where the integrand depends on $\vec{k}$. We marked characteristic points of the geometric setup.  We oriented the axis $k_x$ along $\vec{q}$. As we reduce $q$ , the points "1" and "6" get closer to the $k_x$ axis and merge with analogous points from the lower half plane at value $q = 2\Lambda_F$. The situation is analogous for intermediate momenta with the points "2", "3", "4" and "5". In the last case, $q < \Lambda_F$ only the point "0" survives. \\
The points are described by the intersections of two circles of radius $r \in \{ k_F - \Lambda_F, k_F, k_F + \Lambda_F \}$. It is convenient to proceed in the reference frame shifted by the vector $- \vec{q}$. We introduce a function $u(\phi, r, q)$, which describes the distance from the origin of the coordinate system to a point on the circle of radius $r$, centered at point $( q, 0 )$. 
\begin{align}
    u(\phi, r, q)  = q \cos(q) + \sqrt{r^2 - q^2 \sin^2(\phi)} \; .
    \label{u_def}
\end{align}
Below we define the angles which describes the characteristic points in the new frame of reference. The angles are depicted in the Fig. \ref{angles} and the corresponding expressions are:

\begin{figure}[h]
\includegraphics[width= 0.8 \textwidth]{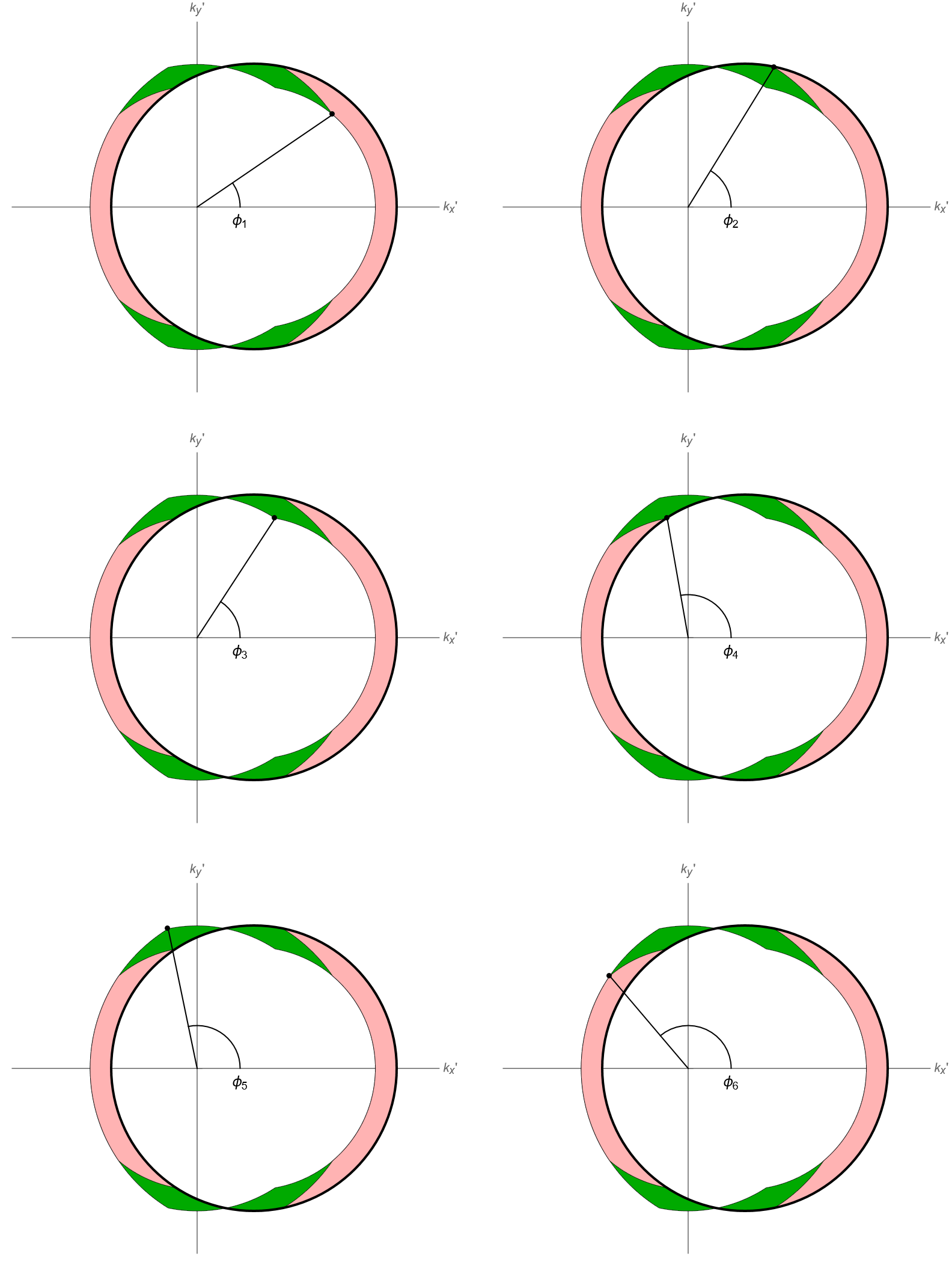}
\caption{ The angles $\phi_1, \phi_2, \phi_3, \phi_4, \phi_5, \phi_6$ -see the text and Eq.(\ref{angles_def}). } 
\label{angles}
\end{figure} 

\begin{align}
    \phi_0 &\coloneqq \arccos \left( \frac{q}{2 k_F} \right) 
    & & \stackrel{ k_F \gg q, \, \Lambda_F }{\longrightarrow} & & \phi_0 \approx \arccos \left( \frac{\pi}{2} \right) & &  \notag \\
    \phi_1 &\coloneqq \arccos \left( \frac{(k_F + \Lambda_F)^2 - (k_F - \Lambda_F)^2 + q^2}{2 q (k_F + \Lambda_F)} \right)
    & & \stackrel{ k_F \gg q, \, \Lambda_F }{\longrightarrow} & & \phi_1 \approx \arccos \left( \frac{2 \Lambda_F}{q} \right)  & & \notag \\
    \phi_2 &\coloneqq \arccos \left( \frac{(k_F + \Lambda_F)^2 - (k_F)^2 + q^2}{2 q (k_F + \Lambda_F)} \right) 
    & & \stackrel{ k_F \gg q, \, \Lambda_F }{\longrightarrow} & & \phi_2 \approx \arccos \left( \frac{ \Lambda_F}{q} \right)  & & \notag \\
    \phi_3 &\coloneqq \arccos \left( \frac{(k_F)^2 - (k_F - \Lambda_F)^2 + q^2}{2 q \, k_F} \right)
    & & \stackrel{ k_F \gg q, \, \Lambda_F }{\longrightarrow} & & \phi_3 \approx \arccos \left( \frac{ \Lambda_F}{q} \right)  
    \label{angles_def}
    & & \\
    \phi_4 &\coloneqq \arccos \left( \frac{(k_F - \Lambda_F)^2 - (k_F )^2 + q^2}{2 q (k_F - \Lambda_F)} \right) 
    & & \stackrel{ k_F \gg q, \, \Lambda_F }{\longrightarrow} & & \phi_4 \approx \arccos \left( - \frac{ \Lambda_F}{q} \right)  & & \notag \\
    \phi_5 &\coloneqq \arccos \left( \frac{(k_F )^2 - (k_F + \Lambda_F)^2 + q^2}{2 q \, k_F} \right) 
    & & \stackrel{ k_F \gg q, \, \Lambda_F }{\longrightarrow} & & \phi_5 \approx \arccos \left( - \frac{ \Lambda_F}{q} \right)  & & \notag \\
    \phi_6 &\coloneqq \arccos \left( \frac{(k_F - \Lambda_F)^2 - (k_F + \Lambda_F)^2 + q^2}{2 q (k_F - \Lambda_F)} \right) 
    & & \stackrel{ k_F \gg q, \, \Lambda_F }{\longrightarrow} & & \phi_6 \approx \arccos \left( - \frac{2 \Lambda_F}{q} \right) \; .  & &  \notag 
\end{align}


\subsection*{ Small bosonic momenta: $\mathcal{X}_{<} \coloneqq \mathcal{X}(|\vec{q}| < \Lambda_F )  $}

In this section we evaluate the expression for $\mathcal{X}_{<}$. For small magnitudes of the bosonic momentum $q < \Lambda_F$ the integrals in Eq.~(\ref{freq_integrated}) simplify because $k_F - \Lambda_F<|\vec{k} + \vec{q}| < k_F + \Lambda_F $:

\begin{align}
    \mathcal{X}_{<}(|\vec{q}|, q_0, \Lambda_F) = 
    - \frac{2 g^2 v_F}{(2 \pi)^2}
    \Biggl\{
        \Bigl(1 + \frac{\Lambda_F}{k_F} \Bigr)
        \int_{k_F}^{k_F + \Lambda_F} {\rm d} k 
        \int_0^{2 \pi} {\rm d}\phi \; k \;
        \Biggl[
            \frac{ \Theta \Bigl[ k_F - |\vec{k} + \vec{q}| \Bigr] }{ \bigl( q_0 + i \, \xi_{k_F + \Lambda_F} - i \, \xi_{k_F - \Lambda_F} \bigr)^2}
            + 
            \left( q_0 \to - q_0 \right) \;
        \Biggr] \; 
        +  \\ \notag
        + \, \Bigl(1 - \frac{\Lambda_F}{k_F} \Bigr)
        \int_{k_F - \Lambda_F}^{k_F } {\rm d} k 
        \int_0^{2 \pi} {\rm d}\phi \; k \;
        \Biggl[
            \frac{ \Theta \Bigl[   |\vec{k} + \vec{q}| - k_F \Bigr] }{ \bigl( q_0 + i \, \xi_{k_F + \Lambda_F} - i \, \xi_{k_F - \Lambda_F} \bigr)^2}
            + 
            \left( q_0 \to - q_0 \right) \;
        \Biggr]
    \Biggr\} \; .
\end{align}
\noindent Further calculations can be done analytically. The integrals represent the area of the restricted region of integration (the green region in right plot in Fig. \ref{geometry}), because of no $\vec{k}$ dependence. We obtain:

\begin{align}
\label{Xsm_full}
    \mathcal{X}_<(|\vec{q}|, q_0, \Lambda_F) =  - \frac{2 g^2 v_F}{ \pi^2}
    \; \frac{q_0^2 - 4 v_F^2 \Lambda_F^2}{\left( q_0^2 + 4 v_F^2 \Lambda_F^2 \right)^2} \;
    \left(
        2 k_F^2 \arcsin(\frac{q}{2 k_F}) +
        \frac{q}{2} \sqrt{4 k_F^2 - q^2}
    \right) \; .
\end{align}
For $k_F \gg q$ and $k_F \gg \Lambda_F$ Eq.(\ref{Xsm_full}) simplifies to:

\begin{align}
    \mathcal{X}_<(|\vec{q}|, q_0, \Lambda_F) \approx  - \mathcal{N}_< \; q
    \; \frac{q_0^2 - 4 v_F^2 \Lambda_F^2}{\left( q_0^2 + 4 v_F^2 \Lambda_F^2 \right)^2} \; ,
    \label{Xsm}
\end{align}

\noindent where $\mathcal{N}_< \coloneqq \frac{4 g^2 v_F k_F}{ \pi^2}$. We now integrate Eq.~(\ref{Xsm}) over the cutoff scale from the upper cutoff $\Lambda_u$ to $\Lambda_F$ and get a contribution to the bosonic self energy:

\begin{align}
    B_<(q, q_0, \Lambda_F, \Lambda_u ) =  - \mathcal{N}_< \; q 
    \left(
        \frac{\Lambda_F}{q_0^2 + 4 v_F^2 \Lambda_F^2} - \frac{\Lambda_u}{q_0^2 + 4 v_F^2 \Lambda_u^2}
    \right) 
    \quad \stackrel{ \Lambda_u \to \infty }{\longrightarrow} \quad
    - \mathcal{N}_< \; q 
    \left(
        \frac{\Lambda_F}{q_0^2 + 4 v_F^2 \Lambda_F^2} 
    \right) 
    \; .
\end{align}


\subsection*{ Intemediate bosonic momenta: $\mathcal{X}_{M} \coloneqq \mathcal{X}( |\vec{q}| \in ( \Lambda_F, 2 \Lambda_F ) )  $ }

Now we restrict to $|\vec{q}| \in (\Lambda_F, 2 \Lambda_F )$. We start from the Eq.(\ref{freq_integrated}). The axis $k_x$ is oriented along $\vec{q}$ and centered in the middle of the Fermi surface. We introduce a new frame of reference: $k_x' = k_x + q_x, k_y' = k_y$ and use the angles defined in Eq.(\ref{angles_def}) and shown in Fig. \ref{angles}. We can write the expression for $\mathcal{X}_M(|\vec{q}|, q_0, \Lambda_F)$ as follows:

\begin{align}
    \mathcal{X}_M(|\vec{q}|, q_0, \Lambda_F) = &
    - \frac{g^2 v_F}{ \pi^2}
    \Biggl\{
        \Bigl(1 + \frac{\Lambda_F}{k_F} \Bigr)
        \int_{\phi_4}^{\pi} {\rm d}\phi \; 
        \int_{u(\phi, k_F, q)}^{k_F - \Lambda_F} {\rm d} k 
        \Biggl[
            \frac{ k }{ \bigl( q_0 + i \, \frac{k^2 - (k_F + \Lambda_F)^2}{2 m_F} \bigr)^2}
            + 
            \left( q_0 \to - q_0 \right) \;
        \Biggr] \; 
        +  \notag \\ 
        &+ \, \Bigl(1 - \frac{\Lambda_F}{k_F} \Bigr)
        \int_0^{\phi_2} {\rm d}\phi \; 
        \int_{k_F + \Lambda_F}^{u(\phi, k_F, q)} {\rm d} k 
        \Biggl[
            \frac{ k }{ \bigl( q_0 + i \, \frac{k^2 - (k_F - \Lambda_F)^2}{2 m_F} \bigr)^2}
            + 
            \left( q_0 \to - q_0 \right) \;
        \Biggr] \; 
        +  \\
        &+ \, \Bigl(1 + \frac{\Lambda_F}{k_F} \Bigr)
        \frac{2 (q_0^2 - 4 v_F^2 \Lambda_F^2)}{(q_0^2 + 4 v_F^2 \Lambda_F^2)^2}
        \left(
            \int_{\phi_0}^{\phi_4} {\rm d}\phi \; 
            \int_{u(\phi, k_F, q)}^{k_F} {\rm d} k  \; k + 
            \int_{\phi_4}^{\phi_5} {\rm d}\phi \; 
            \int_{k_F - \Lambda_F}^{k_F} {\rm d} k  \; k +
            \int_{\phi_5}^{\pi} {\rm d}\phi \; 
            \int_{k_F - \Lambda_F}^{u(\phi, k_F + \Lambda_F, q)} {\rm d} k  \; k 
        \right) \; 
        + \notag  \\ \notag
        &+ \, \Bigl(1 - \frac{\Lambda_F}{k_F} \Bigr)
        \frac{2 (q_0^2 - 4 v_F^2 \Lambda_F^2)}{(q_0^2 + 4 v_F^2 \Lambda_F^2)^2}
        \left(
            \int_0^{\phi_2} {\rm d}\phi \; 
            \int_{u(\phi, k_F - \Lambda_F, q)}^{k_F+\Lambda_F} {\rm d} k  \; k + 
            \int_{\phi_2}^{\phi_3} {\rm d}\phi \; 
            \int_{u(\phi, k_F - \Lambda_F, q)}^{u(\phi, k_F , q)} {\rm d} k  \; k +
            \int_{\phi_3}^{\phi_0} {\rm d}\phi \; 
            \int_{k_F }^{u(\phi, k_F , q)} {\rm d} k  \; k 
        \right)
    \Biggr\} \; .
\end{align}

\noindent Here $u$ and the angles $\phi_i$ are defined in Eq.(\ref{u_def}) and Eq.(\ref{angles_def}) respectively. For $k_F \gg q, \; k_F \gg \Lambda_F$, the result simplifies as follows:

\begin{align}
    \mathcal{X}_M(|\vec{q}|, q_0, \Lambda_F) = 
    - \frac{g^2 v_F}{ \pi^2}
    &\Biggl\{
    \frac{q_0^2 - 4 v_F^2 \Lambda_F^2}{ \big( q_0^2 + 4 v_F^2 \Lambda_F^2 \big)^2 }
        \left[
            4 k_F \Lambda_F ( \pi - \phi_5 + \phi_2 ) + 4 k_F q (1 - \sin \big( \phi_2 ) - \sin ( \phi_5 ) \big) + 2 q \Lambda_F \big( \sin ( \phi_2 ) - \sin ( \phi_5) \big)
        \right] \; +
    \notag \\ 
        &- \int_{\phi_4}^{\pi} {\rm d}\phi \;
        \frac{2 m_F v_F \big( \Lambda_F - q \cos ( \phi ) + \frac{q^2}{2 k_F} \sin^2( \phi )  \big) }{ q_0^2 + v_F^2 \big( \Lambda_F - q \cos ( \phi ) + \frac{q^2}{2 k_F} \sin^2(\phi) \big)^2 } 
        - \int_{0}^{\phi_2} {\rm d}\phi \;
        \frac{2 m_F v_F \big( \Lambda_F + q \cos(\phi) - \frac{q^2}{2 k_F} \sin^2 ( \phi )  \big) }{ q_0^2 + v_F^2 \big( \Lambda_F + q \cos ( \phi) - \frac{q^2}{2 k_F} \sin^2 ( \phi ) \big)^2 } \; +
    \notag \\ 
        & \hspace{8 cm} +\big( \pi - \phi_4 + \phi_2 \big) \frac{4 m_F v_F \Lambda_F}{q_0^2 + 4 v_F^2 \Lambda_F^2} \;
    \Biggr\} \; .
\end{align}

\noindent In fact $\mathcal{X}_M$ does not contribute to the Landau damping, as it is explained below.


\subsection*{ Large bosonic momenta regime: $\mathcal{X}_{>} \coloneqq \mathcal{X}(|\vec{q}| > 2 \Lambda_F ) $}

In this section we address the topic of large bosonic momenta: $|\vec{q}| > 2 \Lambda_F $. Following the same procedure as above we obtain:

\begin{align}
    \mathcal{X}_>(|\vec{q}|, q_0, \Lambda_F) = 
    - \frac{g^2 v_F}{ \pi^2}
    \Biggl\{
        \Bigl(1 + \frac{\Lambda_F}{k_F} \Bigr)
        \Biggl[
            &\int_{\phi_6}^{\pi} {\rm d}\phi \; 
            \int_{u(\phi, k_F, q)}^{u(\phi, k_F + \Lambda_F, q)} {\rm d} k 
            \left( 
                \frac{ k }{ \bigl( q_0 + i \, \frac{(k_F + \Lambda_F)^2 - k^2}{2 m_F} \bigr)^2}
                +
                \left( q_0 \to - q_0 \right) \; 
            \right) +
        \notag \\ \notag
            + &\int_{\phi_4}^{\phi_6} {\rm d}\phi \; 
            \int_{u(\phi, k_F, q)}^{k_F - \Lambda_F} {\rm d} k 
            \left(
                \frac{ k }{ \bigl( q_0 + i \, \frac{(k_F + \Lambda_F)^2 - k^2}{2 m_F} \bigr)^2}
                +
                \left( q_0 \to - q_0 \right)
            \right)
        \Biggr] \; + 
        \\ 
        + \Bigl(1 - \frac{\Lambda_F}{k_F} \Bigr)
        \Biggl[
            &\int_{0}^{\phi_1} {\rm d}\phi \; 
            \int_{u(\phi, k_F - \Lambda_F, q)}^{u(\phi, k_F , q)} {\rm d} k 
            \left( 
                \frac{ k }{ \bigl( q_0 + i \, \frac{k^2 - (k_F - \Lambda_F)^2 }{2 m_F} \bigr)^2}
                +
                \left( q_0 \to - q_0 \right) \; 
            \right) +
        \\ \notag
            + &\int_{\phi_1}^{\phi_2} {\rm d}\phi \; 
            \int_{k_F + \Lambda_F}^{u(\phi, k_F, q)} {\rm d} k 
            \left(
                \frac{ k }{ \bigl( q_0 + i \, \frac{k^2 - (k_F - \Lambda_F)^2}{2 m_F} \bigr)^2}
                +
                \left( q_0 \to - q_0 \right)
            \right)
        \Biggr] \; + 
        \\ \notag
        + \Bigl(1 + \frac{\Lambda_F}{k_F} \Bigr) \;
        &\frac{2 (q_0^2 - 4 v_F^2 \Lambda_F^2)}{(q_0^2 + 4 v_F^2 \Lambda_F^2)^2}
        \left(
            \int_{\phi_0}^{\phi_4} {\rm d}\phi \; 
            \int_{u(\phi, k_F, q)}^{k_F} {\rm d} k  \; k + 
            \int_{\phi_4}^{\phi_5} {\rm d}\phi \; 
            \int_{k_F - \Lambda_F}^{k_F} {\rm d} k  \; k +
            \int_{\phi_5}^{\phi_6} {\rm d}\phi \; 
            \int_{k_F - \Lambda_F}^{u(\phi, k_F + \Lambda_F, q)} {\rm d} k  \; k 
        \right) \; +
        \\ \notag
        + \Bigl(1 - \frac{\Lambda_F}{k_F} \Bigr) \;
        &\frac{2 (q_0^2 - 4 v_F^2 \Lambda_F^2)}{(q_0^2 + 4 v_F^2 \Lambda_F^2)^2}
        \left(
            \int_{\phi_1}^{\phi_2} {\rm d}\phi \; 
            \int_{u(\phi, k_F - \Lambda_F, q)}^{k_F+\Lambda_F} {\rm d} k  \; k + 
            \int_{\phi_2}^{\phi_3} {\rm d}\phi \; 
            \int_{u(\phi, k_F - \Lambda_F, q)}^{u(\phi, k_F , q)} {\rm d} k  \; k +
            \int_{\phi_3}^{\phi_0} {\rm d}\phi \; 
            \int_{k_F }^{u(\phi, k_F , q)} {\rm d} k  \; k 
        \right)
    \Biggr\} \; .
\end{align}

\noindent For $k_F \gg q, \; k_F \gg \Lambda_F$ this reduces to:

\begin{align}
    \mathcal{X}_>(|\vec{q}|, q_0, \Lambda_F) = 
    - \frac{\mathcal{N}_<}{ 2}
    &\Biggl\{
        q \int_{\phi_6}^{\pi} {\rm d}\phi \;
        \frac{ \cos(\phi) - \frac{q}{2 k_F} \sin^2(\phi) }{q_0^2 + v_F^2 q^2 \big( \cos(\phi) - \frac{q}{2 k_F} \sin^2(\phi) \big)^2}
        -  q \int_{\phi_5}^{\pi} {\rm d}\phi \;
        \frac{ \cos(\phi) - \frac{q}{2 k_F} \sin^2(\phi) - \frac{\Lambda_F}{q} }{q_0^2 + v_F^2 q^2 \big( \cos(\phi) - \frac{q}{2 k_F} \sin^2(\phi) -  \frac{\Lambda_F}{q} \big)^2 } + 
    \notag \\ \notag
        &-  q \int_{0}^{\phi_1} {\rm d}\phi \;
        \frac{ \cos(\phi) - \frac{q}{2 k_F} \sin^2(\phi) }{q_0^2 + v_F^2 q^2 \big( \cos(\phi) - \frac{q}{2 k_F} \sin^2(\phi)  \big)^2 } +
        q \int_{0}^{\phi_2} {\rm d}\phi \;
        \frac{ \cos(\phi) - \frac{q}{2 k_F} \sin^2(\phi) + \frac{\Lambda_F}{q} }{q_0^2 + v_F^2 q^2 \big( \cos(\phi) - \frac{q}{2 k_F} \sin^2(\phi) +  \frac{\Lambda_F}{q} \big)^2 } +
    \\ \notag
        &+ \frac{ q_0^2 - 4 v_F^2 \Lambda_F^2}{(q_0^2 + 4 v_F^2 \Lambda_F^2)^2}
        \bigg[
            2 \Lambda_F \big( \phi_2 + \phi_6 - \phi_1 - \phi_5 \big)
            - q \big( 2 \sin(\phi_2) + 2 \sin(\phi_5) - \sin(\phi_1) - \sin(\phi_6) - 2 \sin(\phi_0) \big)
        \bigg] \; +
    \\
        & + \frac{2 \Lambda_F \left( \phi_1 + \phi_5 - \phi_2 - \phi_6 \right) }{q_0^2 + 4 v_F^2 \Lambda_F^2 }
    \Biggr\} \; .
\label{Xbigger_full}
\end{align}


\subsection*{Landau damping}

The aim of the present section is to explain which parts of $\mathcal{X}$ contribute to a non-analytic Landau term after integrating the flow down to $\Lambda_F = 0$. The Landau damping term appears as a result of integrating out all fermions around the Fermi surface and expansion of the bubble diagram around $\frac{|q_0|}{q} \to 0$. For this reason we focus on situation, where $q \gg \Lambda_F$. We find: 

\begin{align}
\label{Bbigger}
    B(q >2 \Lambda_F, q_0, \Lambda_F) = \int_{\infty}^{\Lambda_F} {\rm d} \Lambda' \mathcal{X}(|\vec{q}|, q_0, \Lambda') =  \int_{q/2}^{\Lambda_F} {\rm d} \Lambda' \mathcal{X}_>(|\vec{q}|, q_0, \Lambda') \; + \; \int_{q}^{q/2} {\rm d} \Lambda' \mathcal{X}_M(q, q_0, \Lambda') + \int_{\infty}^{q} {\rm d} \Lambda' \mathcal{X}_<(|\vec{q}|, q_0, \Lambda') \; .
\end{align}

\noindent The last two integrals do not contribute to the Landau damping term, which is clear from equations and can also be easily checked numerically. The entire Landau damping contribution is included in $\mathcal{X}_>$.

\noindent We consider Eq. (\ref{Xbigger_full}) for $q \gg \Lambda_F$, which yields: 

\begin{align}
    \mathcal{X}_>(|\vec{q}|, q_0, \Lambda_F) = 
    - \frac{\mathcal{N}_<}{ 2}
    &\Biggl\{
        q \int_{\phi_6}^{\pi} {\rm d}\phi \;
        \frac{ \cos(\phi) - \frac{q}{2 k_F} \sin^2(\phi) }{q_0^2 + v_F^2 q^2 \big( \cos(\phi) - \frac{q}{2 k_F} \sin^2(\phi) \big)^2}
        -  q \int_{\phi_5}^{\pi} {\rm d}\phi \;
        \frac{ \cos(\phi) - \frac{q}{2 k_F} \sin^2(\phi) - \frac{\Lambda_F}{q} }{q_0^2 + v_F^2 q^2 \big( \cos(\phi) - \frac{q}{2 k_F} \sin^2(\phi) -  \frac{\Lambda_F}{q} \big)^2 } + 
    \notag \\ \notag
        &-  q \int_{0}^{\phi_1} {\rm d}\phi \;
        \frac{ \cos(\phi) - \frac{q}{2 k_F} \sin^2(\phi) }{q_0^2 + v_F^2 q^2 \big( \cos(\phi) - \frac{q}{2 k_F} \sin^2(\phi)  \big)^2 } +
        q \int_{0}^{\phi_2} {\rm d}\phi \;
        \frac{ \cos(\phi) - \frac{q}{2 k_F} \sin^2(\phi) + \frac{\Lambda_F}{q} }{q_0^2 + v_F^2 q^2 \big( \cos(\phi) - \frac{q}{2 k_F} \sin^2(\phi) +  \frac{\Lambda_F}{q} \big)^2 } +
    \\ 
        &+ 2 \frac{\Lambda_F^2}{q} \left( \frac{4 v_F^2 \Lambda_F^2 -  q_0^2 }{(q_0^2 + 4 v_F^2 \Lambda_F^2)^2}
        \; + 
        \frac{2 }{q_0^2 + 4 v_F^2 \Lambda_F^2 } \right)
    \Biggr\} \; .
\label{Xbigger_large_q}
\end{align}

\noindent The first two lines only yield analytical contributions. The integration over $\Lambda$ according to Eq.(\ref{Bbigger}) can be applied to the last line in Eq. (\ref{Xbigger_large_q}) leading to:

\begin{align}
    B_> \approx - \frac{\mathcal{N}_<}{ 4 v_F^3} 
    \left[
        v_F \frac{\Lambda_F}{q} \bigg( 3 + \frac{q_0^2}{q_0^2 + 4 v_F^2 \Lambda_F^2} \bigg) - 
         \frac{v_F}{2} \bigg( 3 + \frac{q_0^2}{q_0^2 + v_F^2 q^2} \bigg) +
         2 \frac{q_0}{q} \left( \arctan\bigg( \frac{2 v_F \Lambda_F}{q_0} \bigg) - \arctan\bigg( \frac{v_F q}{q_0} \bigg) \right)
    \right] \; .
\end{align}

\noindent After taking $\Lambda_F \to 0$ and $q_0/q \to 0$ one obtains the Landau damping term with the correct coefficient. This conclusion was also checked numerically.


\section*{Flow equations}

Below we present the flow equations for the scale-dependent order parameter $\rho_0^{\Lambda} \coloneqq \tfrac{1}{ 2 } (\phi_0^{\Lambda})^2$ and the bosonic self-interaction vertex constant $u^{\Lambda}$.

\begin{align}
    \partial_\Lambda \rho_0^{\Lambda} = \frac{3}{2 (2 \pi)^3} \int_{-\infty}^{\infty} {\rm d} q_0 \int {\rm d}^2 q \frac{\partial_\Lambda R_b}{\left( m_{\Lambda}^2 + Z \big( |\vec{q}| - \Lambda_F(\Lambda) \big)^2 + A q_0^2 + B \big( \vec{q}, q_0, \Lambda_F(\Lambda) \big) + R_b(\vec{q}) \right)^2} \\
    \partial_\Lambda u^{\Lambda} = \frac{9}{(2 \pi)^3} \int_{-\infty}^{\infty} {\rm d} q_0 \int {\rm d}^2 q \frac{\partial_\Lambda R_b}{\left( m_{\Lambda}^2 + Z \big( |\vec{q}| - \Lambda_F(\Lambda) \big)^2 + A q_0^2 + B \big( \vec{q}, q_0, \Lambda_F ( \Lambda) \big) + R_b(\vec{q}) \right)^3} \; .
\end{align}

\noindent For $\Lambda_0 = 0$ we chose the Litim cut-off function as a regulator: 

\begin{align}
    R_b(\vec{q}) = Z \left( \Lambda^2 - (|\vec{q}| - \Lambda_F(\Lambda))^2 \right) \Theta \left[ \Lambda^2 - (|\vec{q}| - \Lambda_F(\Lambda))^2 \right] \; ,
\end{align}

\noindent while for $\Lambda_0 > 0 $ we used:

\begin{align}
    R_b(\vec{q}) = Z \left( \Lambda^2 - |\vec{q}|^2 \right) \Theta \left[ \Lambda^2 - |\vec{q}|^2 \right] \; .
\end{align}

\end{document}